\documentclass[12pt]{iopart}
\usepackage{multirow}
\usepackage{axodraw}
\usepackage{iopams}

\newcommand{\lt}[1]{\overline{#1}#1}

\begin{document}

\title{Breaking of $B-L$ in superstring inspired $E_6$ model.}

\author{A. Kartavtsev}

\address{Institut f\"{u}r Physik, Universit\"{a}t Dortmund, 
D-44221 Dortmund, Germany}

\ead{akartavt@het.physik.uni-dortmund.de}

\begin{abstract}
In the framework of the superstring inspired $E_6$ model, `low--energy' extensions
of the standard model compatible with leptogenesis are considered and masses
of right--handed neutrinos in two scenarios allowed by long--lived protons are
discussed. The presence of two additional generations allows breaking of
$B-L$ without generating nonzero vacuum expectation values of right--handed
sneutrinos of the three known generations. After the symmetry breaking,
right--handed neutrinos acquire Majorana masses $M_{\nu^c}\simeq 10^{11}$ GeV. 
Within the framework of a simple discrete symmetry, assumptions made to provide
a large mass of right--handed neutrinos are shown to be self--consistent.
Supersymmetric structure of the theory ensures that large corrections,
associated with the presence of a (super)heavy gauge field, cancel out.
\end{abstract}

\maketitle

\section{\label{introduction}Introduction}
The observed baryon asymmetry of the universe is one of the most interesting problems of
particle physics and cosmology. As was suggested by Sakharov, the cosmological baryon
excess can be generated through the baryon number violating decays of heavy particles \cite{Sakharov}.
Any microscopic theory able to reproduce the observed asymmetry must satisfy Sakharov's
three conditions:
\begin{itemize}
\item  baryon (or baryon minus lepton) number non-conservation;
\item C and CP violation;
\item deviation from thermal equilibrium.
\end{itemize}

The condition of baryon number violation makes a strong appeal to physics beyond the
standard model (SM). It has been proven to be difficult to generate an excess of baryons
through direct violation of the baryon number. A very attractive scenario of generating lepton
number asymmetry through lepton number violating decays of heavy right--handed neutrinos,
which is later converted into the baryon number by sphalerons \cite{'tHooft:1976up, Kuzmin:1985mm}, 
has been suggested by Fukugita and Yanagida \cite{Fukugita:1986hr}. This scenario requires any viable 
extension of the SM to contain heavy right--handed neutrinos, which, furthermore, are the reason for small masses of
conventional neutrinos.

Augmented with three right--handed neutrinos, the SM itself fails to reproduce the observed
baryon asymmetry of the universe $Y_B$ due to the relatively small CP violation and too weak
phase transition. Electroweak baryogenesis in the minimal supersymmetric standard model
(MSSM) is not yet ruled out, but is close to being so
\cite{Orloff:2003bg}. Since a significant amount of
asymmetry is generated during the electroweak (EW) phase transition, from the practical
point of view it is of interest to consider simple `low--energy' extensions of the SM (i.e.,
models with additional
$U(1)$ or $SU(2)$ gauge symmetry and extended particle content) able
to reproduce the observed $Y_B$ \cite{Joshipura:1999is,
Joshipura:2001ui, Joshipura:2001ya}.  From the theoretical point of view, it is interesting
to investigate lepto- and baryogenesis in grand unification theories (GUT). In this paper, an
attempt to derive `low--energy' extensions of the SM, compatible with the Fukugita--Yanagida
scenario, from a supersymmetric GUT model is made.

The choice of a supersymmetric model is motivated by two reasons. First, scalar
superpartners of right--handed neutrinos can be used to break down $B-L$ symmetry
spontaneously by the Higgs mechanism. Second, supersymmetry (if it is broken only softly,
as is usually assumed) provides cancellation of large quantum corrections associated with the
presence of high scales in the theory.

A minimal gauge group with right--handed neutrino contained in the same representation
as the 15 known fermions is $SO(10)$. An additional representation is required to fit Higgs
fields responsible for breaking of the electroweak symmetry. Following the ideas of grand
unification, it seems more natural to choose a gauge group which at the unification scale
treats all states on an equal footing, i.e., which contains the known states along with
right--handed
neutrino and Higgs doublets in the fundamental representation. This is the case for
the exceptional group $E_6$, one of the most prominent candidates for a unified theory. Its
characteristic features are as follows.

\begin{itemize}
\item It naturally follows from breaking of superstring
$E^{'}_8\bigotimes E_8$ with $N_f \textbf{27}+\delta(\textbf{27}+
{\bf \overline{27}})$ chiral superfields  and \textbf{78} vector superfields
as zero mode spectra.
\item It allows chiral representations and its fundamental representation fits the fifteen known
fermions along with right--handed neutrino and two Higgs doublets.
\item Gauge anomalies are automatically canceled.
\end{itemize}

In  \sref{partcontent}, the particle content of the fundamental representation and possible charge
assignments are discussed. There are six charge assignments, compatible with the SM. Charge
conservation in the processes involving states of different generations requires that the same
charge assignment be used for all generations.

Some of the fields in the adjoint representation may lead to a rapid proton decay.
Constraints implied by a long-lived proton are considered in section
\sref{protdecay}. In addition to
the intermediate gauge groups listed in \cite{Matsuoka:1986vg} 
the $SU(5)\times U^2(1)$ is allowed for two charge
assignments. Nevertheless, Yukawa interactions implied by the residual $SU(5)$ symmetry
make the rapid proton decay mediated by new down-type quarks unavoidable, unless those
are very heavy.

\Sref{Majoranamass} discusses $B-L$ symmetry breaking and masses of the right--handed neutrinos.
An interesting feature of the model under consideration is the presence of additional 
$\delta \bf (27+\overline{27})$ generations, scalar right--handed neutrinos of which may be used to break $B-L$ symmetry
spontaneously. Introduced simple discrete symmetry ensures that $B-L$ is broken at the
scale which is sufficiently high for generating large masses of the right--handed neutrinos and
that right--handed scalar neutrinos of the three known generations do not acquire a vacuum
expectation value (VEV). It also forbids Yukawa couplings which, if present, would induce
large masses of the conventional neutrinos. The supersymmetric structure of the theory
ensures that large quantum corrections to masses of scalars, associated with the presence of
heavy gauge fields, cancel out. After the $B-L$ breaking, the `low--energy' gauge group is 
$SU_C(3)\times SU_L(2)\times U^2(1)$.

Finally, \sref{conclusions} is devoted to some concluding remarks.

\section{\label{partcontent}Particle content and charge assignments}
The ten-dimensional $E_8\times E^{'}_8$ heterotic superstring theory compactifies to the
 $M_4\times \Gamma$ and yields a
low--energy theory with $N=1$ supersymmetry. $\Gamma$ is the
Calabi--Yau manifold with $SU(3)$ holonomy. If $\Gamma$  is simply
connected, then $E_8\times E^{'}_8$ gauge group breaks down to the
$E_6\times E^{'}_8$ subgroup. For a multiply connected manifold the initial gauge group breaks down to
 $G\times E^{'}_8$ where G is a subgroup $E_6$ \cite{Witten:1985xc}. In this scheme, chiral superfields $N_f \textbf{27}+\delta(\textbf{27}+ {\bf
\overline{27}})$ and {\bf 78} vector superfields of $E_6$ emerge as
the zero mode spectra. For a wide class of models $\delta=1$. States
in $\delta(\textbf{27}+ {\bf \overline{27}})$ are denoted by $\chi$
and $\overline{\chi}$ respectively, whereas states in $N_f~{\bf 27}$
are denoted by $\psi$.

The fundamental representation of $E_6$ fits the 15 known states of the SM along with two
Higgs doublets ($H^u$ and $H^d$), a pair of down-type quarks ($D$ and $D^c$) and two SM singlets  $\nu^c$ and $S$.
\begin{table}[h]
\caption{\label{charges}Particle content of $E_6$: fundamental representation.}
\begin{indented}
\item[]
\begin{tabular}{llllllllllll}
\br
\centre{6}{The Standard Model states}&\centre{6}{New states}\\
\ms\crule{6}&\crule{6}\\

$B-L$ & $Y$ & $I_3$ & $Q_{em}$ & $P$ & & $B-L$ & $Y$ & $I_3$ & $Q_{em}$ & $P$&\\

1/3 & 1/3 & 1/2  & 2/3 & $u$ & \multirow{2}{*}{$\biggr\}Q$} &  0 & 1 & 1/2  & 1 & $E^c$& \multirow{2}{*}{$\biggr\}H^u$}\\
1/3 & 1/3 & -1/2  & -1/3 & $d$ & &  0 & 1 & -1/2  & 0 & $N^c$&\\

 -1 & -1 & 1/2  & 0 & $\nu$ & \multirow{2}{*}{$\biggr\}L$} & 0 & -1 & 1/2  & 0 & $N$&\multirow{2}{*}{$\biggr\}H^d$}\\
 -1 & -1 & -1/2  & -1 & $e$ & & 0 & -1 & -1/2  & -1 & $E$&\\

 -1/3 & -4/3 & 0  & -2/3 & $u^c$ & &  -2/3 & -2/3 & 0  & -1/3 & $D$&\\

 -1/3 & 2/3 & 0  & 1/3 & $d^c$ & & 2/3 & 2/3 & 0  & 1/3 & $D^c$&\\

 1 & 2 & 0  & 1 & $e^c$ & & 1 & 0 & 0  & 0 & $\nu^c$&\\

\centre{6}{~}  & 0 & 0 & 0  & 0 & $S$&\\
\br
\end{tabular}
\end{indented}
\end{table}

A systematic study of quantum numbers of states in the fundamental and adjoint
representations is conveniently performed using the Cartan--Weyl method 
\cite{Dynkin, Slansky:1981yr}. In the
framework of this method, each state is represented by its weight. Weights of
{\bf 27} and {\bf 78} of $E_6$ are presented in tables \ref{weights_of_27}
and  \ref{weights_of_78}, respectively. The hypercharge of any state is given by a
scalar product of the hypercharge operator $Y$ and the weight of the state. 
One can easily check that the hypercharge assignment is not unique --- there 
are three assignments which reproduce the SM:
\numparts
\begin{eqnarray}
3Y_1=(1,~\bar{1},~1,~\bar{3},~\bar{1},~0),\\
3Y_2=(1,~\bar{1},~\bar{5},~\bar{3},~\bar{1},~0),\\
3Y_3=(1,~\bar{1},~1,~3,~\bar{1},~0).
\end{eqnarray}
\endnumparts
There are also three $B-L$ operators,
i.e. three assignments, which reproduce $B-L$ quantum numbers of the
SM states:
\numparts
\begin{eqnarray}
3B-L_1=(1,~\bar{1},~\bar{2},~\bar{3},~\bar{1},~0),\\
3B-L_2=(1,~\bar{1},~1,~0,~\bar{1},~0),\\
3B-L_3=(1,~\bar{1},~\bar{2},~0,~\bar{1},~0).
\end{eqnarray}
\endnumparts
Different assignments correspond to different embeddings of states
into subgroups of $E_6$.  Out of nine $Y_i\times B-L_j$ combinations
six are compatible with the SM:
\begin{eqnarray}
{\rm I:~~~ }(Y_1,~B-L_1),\qquad{\rm II: ~~}(Y_1,~B-L_2),\qquad{\rm III: ~~}(Y_2,~B-L_1)\nonumber\\
{\rm IV:  ~}(Y_2,~B-L_3),\qquad{\rm V: ~}(Y_3,~B-L_2),\qquad{\rm VI: ~}(Y_3,~B-L_3)\nonumber
\end{eqnarray}

Since $B-L$ is gauged, the Majorana mass of the right--handed neutrino, which is an
essential ingredient of leptogenesis, is forbidden unless $B-L$ symmetry is broken down.
Present data on neutrino masses as well as theoretical estimates of leptogenesis in other GUT
models favour the $10^{10}-10^{12}$ GeV $\nu^c$ mass
range \cite{Joshipura:2001ya,Plumacher:1998ex}.

Using weights, one can derive the form of trilinear interactions
following the algorithm developed in \cite{Anderson:1999em}.
In the flavor basis
\begin{eqnarray}
\label{superpotential}
\fl W= \lambda^{ijk}_1 u_i^{\rm c}(Q_jH_k^{\rm u})+\lambda^{ijk}_2 d_i^{\rm c}(Q_j
H_k^{\rmd})+\lambda^{ijk}_3
e_i^{\rm c}(L_j H_k^{\rmd})+\lambda^{ijk}_4 S_i(H_j^{\rm u}H_k^{\rmd})+
\lambda^{ijk}_5 S_i D_j D_k^{\rm c}\nonumber\\
\fl\qquad +[\lambda^{ijk}_6 e_i^{\rm c} u_j^{\rm c} D_k+\lambda^{ijk}_7
D_i^{\rm c}(Q_j L_k)+\lambda^{ijk}_8
d_i^{\rm c}\nu_j^{\rm c} D_k]+[\lambda^{ijk}_9 D_i(Q_j Q_k)+
\lambda^{ijk}_{10} D_i^{\rm c} u_j^{\rm c} d_k^{\rm c}]\\
\fl\qquad+\lambda^{ijk}_{11}\nu_i^{\rm c}(L_j H_k^{\rm u})\nonumber
\end{eqnarray}

Using \tref{weights_of_27}, one can check that the corresponding weights add up to zero at each vertex.
The form of the superpotential is independent of the charge assignment in the sense that the
change of assignment will only result in `permutation' of vertices. See assignments I, II and
III, for example:

\begin{center}
\begin{indented}
\item[]
\begin{tabular}{llllllll}
\br
& [$10\bar{1}001$] & [$00010\bar{1}$] & [$\bar{1}01\bar{1}00$] &
& [$1\bar{1}01\bar{1}0$] & [$01\bar{1}010$] & [$101\bar{1}00$]\\
\mr
I  &  $\nu^c$  & $\nu$  & $N^c$ &  & $S$ & $N$ & $N^c$  \\
II &  $S$  & $N$  & $N^c$ &  & $\nu^c$ & $\nu$ & $N^c$  \\
III  &  $e^c$  & $\nu$  & $E$ &  & $S$ & $E^c$ & $E$  \\
\br
\end{tabular}
\end{indented}
\end{center}

Along with vertices with $i=j=k$ (i.e. states of the same
generation) superpotential (\ref{superpotential}) necessarily contains terms with the states of 
different generations in one vertex. If the charge
assignment for one of the generations differs from that for other generations, then this results
in nonzero sum of charges in a vertex. For instance, if in the example above assignment II is
used for the first weight and assignment I is used for the second and the third weights then
\begin{eqnarray}
\nu^c\rightarrow S \Rightarrow \nu^c\left (\nu N^c\right)
\rightarrow S \left (\nu N^c\right),&\sum B-L =-1\nonumber\\
S\rightarrow \nu^c \Rightarrow S \left (N N^c\right)
\rightarrow \nu^c \left (N N^c\right),&\sum B-L =1\nonumber
\end{eqnarray}

If instead of assignment II, assignment III is used for the first weight, then it is a sum of
electric charges, which is nonzero. This suggests that the same charge assignment should be
used for all generations.

\section{\label{protdecay}Gauge mediated proton decay.}
Since $SU(5)$ is a subgroup of $E_6$, the gauge sector
of the model contains $X$ and $Y$ bosons, which are known to mediate
proton decay. There are also new gauge fields leading to a rapid
proton decay (\tref{gaugemed}).
\begin{table}[h]
\caption{\label{gaugemed}Gauge fields which mediate proton decay.}
\begin{indented}
\item[]
\begin{tabular}{lllllllllll}
\br
$B-L$ & $Y$ & $I_3$ & $Q_{em}$ & $P$ & $B-L$ & $Y$ & $I_3$ & $Q_{em}$ & $P$\\
\crule{5}&\crule{5}\\
2/3 & 5/3 & 1/2  & 4/3 & $X$ &  -2/3 & 1/3 & 1/2  & 2/3 & $u_{\bar{2}}$\\
2/3 & 5/3 & -1/2  & 1/3 & $Y$ &  -2/3 & 1/3 & -1/2  & -1/3 & $d_{\bar{2}}$\\
\br
\end{tabular}
\end{indented}
\end{table}

To assure that the proton is long--lived, these fields must be
very heavy --- of order of $10^{15}$  GeV or more. If the only source of
masses of those particles is the VEV of neutral scalars, then the masses of 
$(u_{\bar{2}},d_{\bar{2}})$ gauge bosons are determined by
$\langle\tilde{\nu}^c\rangle$, $Y$ boson mass is of
the order of EW symmetry breaking scale and the $X$ bosons remain massless even after all
neutral scalars develop a nonzero VEV.

Consequently, the gauge superfields which mediate proton decay have to become massive
at the first stage of symmetry breaking. If the manifold $\Gamma$
is multiply connected, then the
effective Higgs mechanism \cite{Hosotani:1983xw,
Matsuoka:1985eh} breaks symmetry at the compactification scale and induces
large masses of the gauge fields.

From the discussion above it follows that after $E_8\times E^{'}_8$
breaking the gauge group of
the model is not $E_6$ itself, but a
subgroup $G$ of $E_6$. A very elaborate analysis of many possible breaking
chains has been performed in \cite{Matsuoka:1986vg, Matsuoka:1985eh}.
It was argued there that a gauge
field gets mass of the order of
$O(10^{18})$ GeV if $(Z,\rho)\neq 0$. Here $Z$ is the zero root
breaking $SU_C(3)\times SU_L(2)\times U_Y(1)$ preserving direction,
and $\rho$ is the weight of the gauge field. Scalar products $(Z,\rho)$
for fundamental and adjoint representations of $E_6$ are listed, respectively,
in tabels \ref{weights_of_27} and \ref{weights_of_78}.

\begin{table}[h]
\caption{\label{SU5adjoint}(Z,$\rho$) products for
(\textbf{3},\textbf{2})$_{\bar{5}}$ gauge fields.}
\begin{indented}
\item[]
\begin{tabular}{lllllllllll}
\br
 SU(5)& $SU(3)\times SU(2)$ {\tiny $\times U(1)$} & I & II & III & IV & V & VI
& (Z,$\rho$) \\
\mr
 \textbf{24} & (\textbf{3},\textbf{2})$_{,\bar{5}}$ &  $X$ &  $X$ & $d_{\bar{2}}$ & $d$ & $d_{\bar{2}}$
& $d$ & $\alpha$\\
\textbf{24}& (\textbf{3},\textbf{2})$_{,\bar{5}}$ &  $Y$ &  $Y$ &
$u_{\bar{2}}$ & $u$ & $u_{\bar{2}}$
& $u$ & $\alpha$\\
\br
\end{tabular}
\end{indented}
\end{table}

The requirement of $(Z,\rho)\neq 0$ for the gauge fields which
mediate proton decay does not allow $G=SO(10)\times U(1)$.
(\textbf{3},\textbf{2})$_{\bar{5}}$ states (and their charge
conjugate) are the only states in the adjoint of $SU(5)$ which are
not automatically massless. As seen from \tref{SU5adjoint}, 
the requirement that these fields be massless (i.e. that $SU(5)$ is
unbroken) does not lead to
a rapid gauge mediated proton decay for the charge assignments
 IV and VI. In other words, in addition to the
intermediate gauge groups listed in \cite{Matsuoka:1986vg} as
allowed, $G=SU(5)\times U^2(1)$ is allowed for the charge
assignments IV and VI. Nevertheless, in this case the
residual $SU(5)$ symmetry implies that couplings in
(\ref{superpotential}) are related in the following way;
\[
\lambda_1=\lambda_2=\lambda_4=\lambda_{10},\qquad
\lambda_5=\lambda_7=\lambda_9, \qquad\lambda_3=\lambda_6,
\qquad\lambda_8=\lambda_{11}
\]
and the rapid proton decay mediated by heavy down-type quarks  $D$ and
$D^{\rm c}$ is unavoidable.

Apart from scenarios with an extended color group there are only two options left:
\begin{eqnarray}
\label{u1cube}
G_{U(1)}& =SU_C(3)\times SU_L(2)\times U(1)\times U(1)\times U(1)\\
\label{su2lsu2r}
G_{SU_R(2)} & =SU_C(3)\times SU_L(2)\times SU_R(2)\times U(1)\times U(1)
\end{eqnarray}

If the intermediate scale symmetry group is given by (\ref{u1cube}), then the gauge sector of the SM
is supplemented by two neutral states (the rest are superheavy), denoted by
 $\phi^0$ and $\omega^0$ (see \tref{weights_of_78}).

If the intermediate scale symmetry group is given by (\ref{su2lsu2r}), then the gauge sector of the
SM is supplemented by one $SU_R(2)$ singlet $\omega^0$
and one $SU_R(2)$ triplet $(\phi^c,\phi^0,\overline{\phi}^c)$.
There are three possible $SU_R(3)\rightarrow SU_R(2),_Y$ projections which
result in one left--right symmetric ($R$) and two skew left--right symmetric
($R^{'}$ and $R^{''}$) models. 
$SU_R(2)$ counterparts of $\phi^0$ for different choices of the projection and the charge 
assignment are given in \tref{masslessadj}.
\begin{table}[ht]
\caption{\label{masslessadj}$SU_R(2)$ counterparts of $\phi^0$. Isospin and
hypercharge (as well as $B-L$, unless otherwise indicated) of the states
designated like the SM states are the same as charges of their SM counterparts.
If present, subscript is trice the $B-L$ charge.}
\begin{indented}
\item[]
\begin{tabular}{lllllll}
\br
& \centre{6}{Charge assignment}\\

 & I & II & III & IV & V & VI\\
\mr
$R$ & $e^c_0$,\,$\overline{e}^c_0$ & $e^c$,\,$\overline{e}^c$ &
$e^c_0$,\,$\overline{e}^c_0$ & $e^c$,\,$\overline{e}^c$ &
$\nu^c$,\,$\overline{\nu}^c$ & $\nu^c$,\,$\overline{\nu}^c$ \\

$R^{'}$ & $e^c$,\,$\overline{e}^c$ & $e^c_0$,\,$\overline{e}^c_0$&
$\nu^c$,\,$\overline{\nu}^c$ & $\nu^c$,\,$\overline{\nu}^c$ &
$e^c_0$,\,$\overline{e}^c_0$ & $e^c$,\,$\overline{e}^c$ \\

$R^{''}$ & $\nu^c$,\,$\overline{\nu}^c$ & $\nu^c$,\,$\overline{\nu}^c$ &
$e^c$,\,$\overline{e}^c$ & $e^c_0$,\,$\overline{e}^c_0$ &
$e^c$,\,$\overline{e}^c$ & $e^c_0$,\,$\overline{e}^c_0$\\
\br
\end{tabular}
\end{indented}
\end{table}

While $SU_L(2)$ in (\ref{su2lsu2r}) coincide with that of the standard model,
$U_{Y_L}$ is not the SM $U_{Y}(1)$. Hypercharge $Y$, as well as $B-L$, is
a linear combination of $Q_{Y_L}$, $Q_{Y_R}$ and $I_{3R}$.
An explicit form of gauge interactions
$g_{\alpha}\Lambda_{\alpha}[\psi^+T^{\alpha}\psi]$ for the first charge
assignment and left--right symmetric model can be read off from
\tref{weights_of_27}:
\begin{eqnarray}
\label{gaugeint}
\fl\frac{g_{Y_L}}2\, Y_L \left[\lt{Q}-2\lt{D}-(\lt{H^u}+\lt{H^d})-\lt{L}+
2(\lt{e^c}+\lt{\nu^c})+2\lt{S}\right]\\
+\frac{g_{Y_R}}2\, Y_R \left[-(\lt{d^c}+\lt{u^c})+2\lt{D^c}+(\lt{H^u}+
\lt{H^d})\right.\nonumber\\
\left.-2\lt{L}+(\lt{e^c}+\lt{\nu^c})-2\lt{S}\right]
+\frac{g_{W_R}}2\, W^i_R \left[(d^c,u^c)^+\tau_i(d^c,u^c)\right.\nonumber\\
\left.+(H^u,H^d)^+\tau_i (H^u,H^d)+(e^c,\nu^c)^+\tau_i (e^c,\nu^c)\right]+\dots
\nonumber
\end{eqnarray}
At the unification scale the relation among the gauge couplings is as follows:
\begin{equation}
g_{W_R}=\sqrt{3}\, g_{Y_L}=\sqrt{3}\, g_{Y_R}=g_{E_6}
\label{unification}
\end{equation}
Instead of $Y_L$, $Y_R$ and $W^0_R$ one can use their linear combination.
For instance, in the limit (\ref{unification}) linear combination
$Y_{B-L}=\frac1{\sqrt{2}}(Y_L+Y_R)$ is a gauge field of $U_{B-L}(1)$. For the
later purposes it is useful to choose the linear combinations $\omega^{'}_0$,
$\phi^{'}_0$, $\gamma^{'}_0$ in such a way, that only one of the new fields
interacts with right--handed neutrinos and one of the fields interacts with
neither of the SM singlets. After the Standard Model singlets develop VEVs,
$\omega^{'}_0$ and $\phi^{'}_0$ acquire masses, while $\gamma^{'}_0$, which
corresponds to $U_Y(1)$, remains massless. Rewritten in terms of these fields,
(\ref{gaugeint}) takes the form
\begin{eqnarray}
\label{physgaugeint} \fl g_{\gamma^{'}_0}\, \gamma^{'}_0\, \frac16
\left[\lt{Q}-4\lt{u^c}+2\lt{d^c}+
2\lt{D^c}-2\lt{D}+3\lt{H^u}-3\lt{H^d}-3\lt{L}+6\lt{e^c}\right]\\
+g_{\phi^{'}_0}\, \phi^{'}_0\, \left[-\lt{Q}-\lt{u^c}-2\lt{d^c}+3\lt{D^c}+
2\lt{D}+2\lt{H^u}+3\lt{H^d}\right.\nonumber\\
\left.-2\lt{L}-\lt{e^c}-5\lt{S}\right]
+g_{\omega^{'}_0}\, \omega^{'}_0\, \left[\lt{Q}+\lt{u^c}-2\lt{d^c}+\lt{D^c}-
2\lt{D}\right.\nonumber\\
\left.-2\lt{H^u}+\lt{H^d}-2\lt{L}+\lt{e^c}+4\lt{\nu^c}+\lt{S}\right]
\nonumber+\dots
\end{eqnarray}
At the unification scale gauge couplings $g_{\gamma^{'}_0}$,
$g_{\phi^{'}_0}$, $g_{\omega^{'}_0}$ are given by
\begin{equation}
g_{\gamma^{'}_0}=\sqrt{\frac35}\,g_{E_6},~
g_{\phi^{'}_0}=\frac{g_{E_6}}{\sqrt{40}},~
g_{\omega^{'}_0}=\frac{g_{E_6}}{\sqrt{24}}
\end{equation}
Reexpessed in terms of $\omega^{'}_0$, $\phi^{'}_0$, $\gamma^{'}_0$ gauge
interactions of neutral fields are obviously given by
(\ref{physgaugeint}) irrespectively of the intermediate scale symmetry group
$G$, charge assignment or particular choice of $SU_R(3)\rightarrow SU_R(2),\,_Y$
projection.

\section{\label{Majoranamass}Large Majorana masses of the right--handed neutrinos.}
There are two SM singlets whose scalar superpartners may be used to
break symmetry down to the SM: $S$ and the right--handed neutrino.

The former has zero $B-L$ charge, while the later has $B-L=1$. Therefore, it
is the VEV of the scalar superpartner of the right--handed neutrino that breaks $B-L$
symmetry. $S$ couples to Higgs doublets and its VEV $\langle\tilde{S}\rangle$
is the origin of the $\mu$--term: $\mu=\lambda_4 \langle\tilde{S}\rangle$.

If the right--handed sneutrino which develops the VEV couples to states of the three known
generations, it induces huge Dirac masses for components of $L$
and $H^u$ doublets via the last term in (\ref{superpotential}). Neglecting
the possibility that one of the $\nu^c$ superfields is decoupled from the
other states of the three known generations, one comes to the conclusion that all
$\langle\tilde{\nu^c}\rangle=0$, and the $B-L$ symmetry is broken
spontaneously by nonzero $\langle\tilde{\chi}_{\nu^c}\rangle$ and
$\langle\tilde{\overline{\chi}}_{\nu^c}\rangle$ and, consequently,
that $\chi_{\nu^c}$ and $\overline{\chi}_{\nu^c}$ are zero modes.

According to \cite{Matsuoka:1985eh}  chiral superfields in $\delta
({\bf 27}+{\bf\overline{27}})$ can be massive through the Yukawa
coupling ${\bf 27\cdot \overline{27}\cdot 78}$. If $(Z,\rho)\neq 0$
for a component of {\bf 27} or ${\bf \overline{27}}$ with weight
$\rho$, the corresponding chiral superfield gets compactification scale
mass, while $N_f~ {\bf 27}$ chiral superfields remain massless.

\Tref{weights_of_27} shows, that for both discussed
intermediate gauge groups $G_{U(1)}$ and $G_{SU_R(2)}$ and any
charge assignment, it is possible to have massless $\chi_{\nu^c}$ and
$\overline{\chi}_{\nu^c}$. In the case of $G_{U(1)}$ right--handed neutrinos
$\chi_{\nu^c}$ and $\overline{\chi}_{\nu^c}$ are the only massless states
in $\delta ({\bf 27}+{\bf \overline{27}})$.

In the case of $G_{SU_R(2)}$ the number of zero modes in $\delta ({\bf
27}+{\bf \overline{27}})$ depends on the charge
assignment and the particular choice of $SU_R(3)\rightarrow SU_R(2)$ projection (\tref{masslessfund}).
\begin{table}[h]
\begin{flushright}
\caption{\label{masslessfund}States in $\delta ({\bf
27}+{\bf\overline{27}})$ which remain massless after the
compactification. States of {\bf 27} and ${\bf\overline{27}}$ are
labeled here by the same symbol. For instance, $\nu^c$ stands for
both $\chi_{\nu^c}$ and $\overline{\chi}_{\nu^c}$. Components of
$SU_R(2)$ doublets are put into brackets.} \footnotesize
\begin{tabular}{lllllll}
\br
& \centre{6}{Charge assignment}\\

& I & II & III & IV & V & VI\\
\mr $R$ & $(\nu^c,e^c)$ & $\nu^c$, $(H^u,L)$ & $(\nu^c,e^c)$ &
$\nu^c$,
$(H^u,L)$ & $(\nu^c,S)$ & $(\nu^c,S)$\\
$R^{'}$ & $\nu^c$, $(H^u,L)$  & $(\nu^c,e^c)$ & $(\nu^c,S)$ &
$(\nu^c,S)$  & $(\nu^c,e^c)$  & $\nu^c$,
$(H^u,L)$ \\
$R^{''}$ & $(\nu^c,S)$  & $(\nu^c,S)$ & $\nu^c$, $(H^u,L)$ &
$(\nu^c,e^c)$   &
$\nu^c$, $(H^u,L)$ & $(\nu^c,e^c)$  \\
\br
\end{tabular}
\end{flushright}
\end{table}

If supersymmetry is exact, there are no negative mass squared terms
needed to break $U_{B-L}(1)$ down spontaneously by Higgs mechanism.
\begin{equation}
\begin{array}{l}
\displaystyle m^2_{\chi}|\tilde{\chi}_{\nu^c}|^2+
m^2_{\overline{\chi}}|\tilde{\overline{\chi}}_{\nu^c}|^2-
m^2_{\nu^c\,ij}\tilde{\nu}^{c*}_i\tilde{\nu}^{c}_j, \qquad
|m^2_{\overline{\chi}}|, |m^2_{\chi}|, |m^2_{\nu^c\,ij}| \sim
m^2_{soft}
\end{array}
\label{NegMassTerms}
\end{equation}
These terms are assumed to come from the $E^{~'}_8$ sector, where supersymmetry
is considered to break down spontaneously. In the gravity-mediated supersymmetry breaking scenario
the magnitude of the soft terms in the visible sector should be roughly of the order of $\displaystyle
m_{soft}\sim \langle F \rangle / M_{Pl}$. For the commonly accepted
value $m_{soft}\sim 10^3$ GeV the scale of
supersymmetry breaking in the hidden sector $\langle F \rangle^\frac12$ is about $10^{11}$ GeV. It is interesting to
note that this value is of the same order as the desired mass scale of the right--handed neutrinos.
One can consider this as a hint that at the stage when right--handed neutrinos acquire masses,
the temperature is still high enough to produce them thermally.

\paragraph{$\bf SU_C(3)\times SU_L(2)\times U^3(1)$ model.}
As is well known, the scalar potential consists of an $F$-term and $D$-term
coming from the chiral superfield trilinear couplings and the gauge
interactions respectively. The renormalizable superpotential
(\ref{superpotential}) does not contain terms relevant for the
symmetry breaking. The scalar potential coming from
gauge interactions (\ref{physgaugeint}) and soft supersymmetry breaking
is of the form
\begin{equation}
\fl V=\frac{g_{\omega^{'}_0}^2}{2} \left[ \tilde{\psi}^{*}\,
T_{\omega^{'}_0}\, \tilde{\psi}+ q_{\nu^c}\tilde{\chi}_{\nu^c}^{*}
\tilde{\chi}_{\nu^c}- q_{\nu^c}\tilde{\overline{\chi}}_{\nu^c}^{*}
\tilde{\overline{\chi}}_{\nu^c} \right]^2 -
\left[m^2_{\chi}|\tilde{\chi}_{\nu^c}|^2+
m^2_{\overline{\chi}}|\tilde{\overline{\chi}}_{\nu^c}|^2\right]
+m^2_{\nu^c\,ij}\tilde{\nu}^{c*}_i\tilde{\nu}^c_j \label{DtermPot}
\end{equation}

On the one hand the VEVs of right--handed sneutrinos of $\delta ({\bf
27}+{\bf \overline{27}})$ are expected to be at least of the same
order as masses of $\nu^c$, i.e. $\langle \tilde{\chi}_{\nu^c}
\rangle, \langle \tilde{\overline{\chi}}_{\nu^c} \rangle \geq
10^{11}$ GeV; on the other hand, such a huge VEV should not generate
large masses of scalar superpartners via the first term in
(\ref{DtermPot}). Consequently, symmetry breaking should occur in the
$D$--flat direction $\langle \tilde{\chi}_{\nu^c} \rangle=\langle
\tilde{\overline{\chi}}_{\nu^c} \rangle$.

Combined with the requirement that all $\langle \tilde{\nu}^c
\rangle$ be zero, this means that contribution of the first term in
(\ref{DtermPot}) vanishes. To have symmetry breaking by the Higgs mechanism in this
direction, the sum of the mass parameters in the second term should be positive:
$
m^2_{\chi}+m^2_{\overline{\chi}}>0 \label{m_chi_condition}
$

Non-renormalizable terms arise due to the interactions with exchange of superheavy fields,
which correspond to excitations of internal degrees of freedom \cite{Dine:1985vv}. The general form of the
non--renormalizable superpotential is
\begin{eqnarray}
\label{NRsuppot} W=&M^{-1}_c\left[a^{ij}_1\,
\nu_i^c\nu_j^c\overline{\chi}_{\nu^c}\overline{\chi}_{\nu^c}+
a^i_2\,
\nu_i^c\chi_{\nu^c}\overline{\chi}_{\nu^c}\overline{\chi}_{\nu^c}+a_3\,
\chi_{\nu^c}\chi_{\nu^c}\overline{\chi}_{\nu^c}\overline{\chi}_{\nu^c}\, \right]
\nonumber\\
&+M^{-3}_c\left[b_1\,
\chi_{\nu^c}\chi_{\nu^c}\chi_{\nu^c}\overline{\chi}_{\nu^c}
\overline{\chi}_{\nu^c}\overline{\chi}_{\nu^c}+b^i_2\,
\nu_i^c\chi_{\nu^c}\chi_{\nu^c}\overline{\chi}_{\nu^c}
\overline{\chi}_{\nu^c}\overline{\chi}_{\nu^c}\right.\\
&\left.+b^{ij}_3\,
\nu_i^c\nu_j^c\chi_{\nu^c}\overline{\chi}_{\nu^c}\overline{\chi}_{\nu^c}\overline{\chi}_{\nu^c}+b^{ijk}_4\,
\nu_i^c\nu_j^c\nu_k^c\overline{\chi}_{\nu^c}\overline{\chi}_{\nu^c}\overline{\chi}_{\nu^c}\,
\right]+\dots\nonumber
\end{eqnarray}
Given that all $\langle \tilde{\nu}^c \rangle$ are zero, only
$M_c^{3-2n}\left(\chi_{\nu^c}\overline{\chi}_{\nu^c}\right)^n$ terms
in (\ref{NRsuppot}) are relevant for the analysis of symmetry
breaking. These terms are invariant with respect to
$\chi_{\nu^c}\leftrightarrow\overline{\chi}_{\nu^c}$ transformation,
whereas soft supersymmetry breaking terms in (\ref{DtermPot}) are
not, unless $m^2_{\chi}=m^2_{\overline{\chi}}$. If this condition is
not satisfied, then $\langle \tilde{\chi}_{\nu^c} \rangle$ can not
be equal to $\langle \tilde{\overline{\chi}}_{\nu^c} \rangle$.
Nevertheless, since gauge coupling $g$ is
of the order of unity, while
$m_{soft}/M_c$ is of many orders of magnitude smaller than unity, the
deviation from the $D$--flat direction $\langle \tilde{\chi}_{\nu^c}
\rangle=\langle \tilde{\overline{\chi}}_{\nu^c} \rangle$ is very
small. Considering $\langle
\tilde{\overline{\chi}}_{\nu^c}\rangle-\langle
\tilde{\chi}_{\nu^c}\rangle$ as a small perturbation in
(\ref{classicalpot}) one finds
\begin{equation}
\langle \tilde{\overline{\chi}}_{\nu^c}\rangle-
\langle \tilde{\chi}_{\nu^c}\rangle\simeq
\frac{m^2_{\overline{\chi}}-m^2_{\chi}}{8\,g^2\,v_0}
\end{equation}
This effect can be entirely neglected in the analysis of
non--renormalizable superpotential (\ref{NRsuppot}). As for gauge interactions, such a deviation from the 
$D$--flat direction will result in the
generation of masses of scalars $\sim m^2_{\overline{\chi}}-m^2_{\chi}$
via the first term in (\ref{DtermPot}) which add to mass terms
coming from the soft supersymmetry breaking. To simplify the analysis $m^2_{\chi}$
and $m^2_{\overline{\chi}}$ are taken to be equal in what follows.

If the first non-vanishing terms in (\ref{NRsuppot}) are
$
M_c^{-1}(\nu^c\overline{\chi}_{\nu^c})^2 +
M_c^{3-2n}(\chi_{\nu^c}\overline{\chi}_{\nu^c})^n
$
then generated VEV and masses of right--handed neutrinos $\nu^c$ are
\begin{equation}
\langle \tilde{\chi}_{\nu^c} \rangle=\langle
\tilde{\overline{\chi}}_{\nu^c} \rangle\sim
(m_{soft}\,M_c^{2n-3})^{\frac1{2n-2}},\qquad
M_{\nu^c}\sim (m_{soft}\, M_c^{n-2})^{\frac1{n-1}}
\end{equation}
so that $M_{\nu^c}$ is of the order of $10^{3}$ GeV for $n=2$ and is of the order of $10^{11}$ GeV for $n=3$.

Large masses of right--handed neutrinos suggest that $n=3$ and, consequently,
$a_3=0$ in (\ref{NRsuppot}).
Moreover, $a^i_2$ and $b^i_2$, as well as coefficient $c^i_2$ of the  (not
indicated) next similar
term in (\ref{NRsuppot}), are zero to avoid nonzero $\langle \tilde{\nu}^c
\rangle$ and  Dirac--type mass for $\nu^c$. The discrete symmetry of
the compactified manifold possibly accomplishes these conditions
\cite{Lutken:1988zj} as well as the absence of bare mass terms
$M\chi_{\nu^c}\,\overline{\chi}_{\nu^c}$ and $M\nu^c\,\overline{\chi}_{\nu^c}$ in the
superpotential (\ref{superpotential}). There is no reason
to expect scale M to be below compactification scale $M_c$ so that the presence of the first term
would make spontaneous breaking of $B-L$ by
$\chi_{\nu^c}$ and $\overline{\chi}_{\nu^c}$ impossible, while the presence
of the second makes $\nu^c$ a component of the super heavy Dirac neutrino. Just as there are no
$M\chi_{\nu^c}\,\overline{\chi}_{\nu^c}$ and $M\nu^c\,\overline{\chi}_{\nu^c}$
terms, there are no soft supersymmetry breaking terms
$b_{\chi\overline{\chi}}\,\tilde{\chi}_{\nu^c}\,\tilde{\overline{\chi}}_{\nu^c}$
and
$b_{\nu^c\overline{\chi}}\,\tilde{\nu^c}\,\tilde{\overline{\chi}}_{\nu^c}$ in
(\ref{DtermPot}).
If present, the contributions of $b_3$ and $b_4$ terms
are small and neglected in the following discussion. Then the classical potential is of the form
\begin{eqnarray}
\label{classicalpot}
V=& 4 M_c^{-2}\,u^2 \varrho_i\varrho_j
\left[a_{li}a^{*}_{lj} u^2+a_{ij}a^{*}_{nm} \varrho_n\varrho_m\right]+
9 b_1 b_1^{*}M_c^{-6}u^4 v^4\left[u^2+v^2\right]\\
&+6 M_c^{-4}\left[b_1^{*}a_{ij}+b_1a^{*}_{ij}\right]\varrho_i\varrho_j
v^3 u^3 +\frac{g^2 q_{\nu^c}^2}2
\left[\varrho_i\varrho_i+v^2-u^2\right]^2\nonumber\\
&-\left[m^2_{\chi}v^2+m^2_{\overline{\chi}}u^2\right]+
m^2_{\nu^c\,ij}\varrho_i\varrho_j\nonumber
\end{eqnarray}
with $\varrho_i=\langle\tilde{\nu}^c_i \rangle$,
$v=\langle \tilde{\chi}_{\nu^c} \rangle$,
$u=\langle \tilde{\overline{\chi}}_{\nu^c}\rangle$.
For zero $\varrho_i$ the D--flat direction  is defined by $u^2=v^2$ and
the minimum of the potential (\ref{classicalpot}) corresponds to
\begin{equation}
\label{vev&mass}
v_0=\sqrt[8\,]{M_c^6(m^2_{\chi}+m^2_{\overline{\chi}})/(90|b_1|^2)},\qquad
M_{\nu_c}^{ij}=2a_1^{ij}v^2_0 M_c^{-1}
\end{equation}
For nonzero $\varrho_i$ D--flat direction  is defined by
$u^2=v^2+\varrho_i\varrho_i$. The set of products $\varrho_i\varrho_j$
being considered as parameters, the minimizing of (\ref{classicalpot})
with respect to $v$ gives $v$ as function of $\varrho_i\varrho_j$.
True vacuum corresponds to the set $\varrho_i\varrho_j$ which
minimizes $V(v(\varrho_i\varrho_j),\varrho_i\varrho_j)$. Expanding
in powers of $\varrho_i\varrho_j$  in the vicinity of $\varrho_i\varrho_j=0$
and having in mind that partial derivative with respect to $v$ is zero:
\begin{eqnarray}
\label{expansion}
V(v(\varrho_i\varrho_j),\varrho_i\varrho_j)\simeq V(v_0,0)
+\left(m^{2~ij}_{\nu^c}-m^2_{\overline{\chi}}\,\delta^{ij}\right)
\varrho_i\varrho_j\\
+ v_0^4 M_c^{-2}\left[4 a_1^{li}a_1^{*lj}+
6(b_1^{*}a_1^{ij}+b_1a_1^{*ij}) v_0^2 M_c^{-2}+45b_1
b_1^{*} v_0^4 M_c^{-4}\delta^{ij}\right]\varrho_i\varrho_j\nonumber
\end{eqnarray}

Since $v_0/M_c$ ratio is small while $v^2_0/M_c$ is large, the derivative
(\ref{expansion}) is dominated by the first term in square brackets
which is a positive definite. Therefore, $\varrho_i=0$ is at
least a local minimum of the potential (\ref{classicalpot}) for a wide
range of parameters.

For large $\varrho_i\varrho_j$ it is sufficient to
keep only the first term in (\ref{classicalpot}), and an explicit calculation
shows that  $V(v(\varrho_i\varrho_j),\varrho_i\varrho_j)$ grows with growing
$\varrho_i\varrho_j$, i.e. $\varrho_i\varrho_j=0$ is a global minimum of the
classical potential:
\begin{equation}
\label{largevarrho}
\fl V(v(\varrho_i\varrho_j),\varrho_i\varrho_j)=
-M_c^2 (m^2_{\tilde{\chi}_{\nu^c}}+m^2_{\tilde{\overline{\chi}}_{\nu^c}})^2
(16 a_1^{li}a_1^{*lj}\varrho_i\varrho_j)^{-1}+
(m^2_{\tilde{\chi}_{\nu^c}}\delta^{ij}+m^{2\,ij}_{\nu^c})\varrho_i\varrho_j
\end{equation}

A discrete symmetry which allows nonzero $a_1^{ij}$, $b_1$ and forbids
nonzero $a^{i}_2$, $a_3$, $b_2$ couplings is essential for having
large Majorana masses for right--handed neutrinos after symmetry breaking.
Suppose that  right--handed neutrinos
$\nu^c$, $\chi_{\nu^c}$, $\overline{\chi}_{\nu^c}$
acquire additional phases under transformations of the discrete symmetry:
\begin{equation}
\label{transformations}
\nu^c\rightarrow \nu^c\,e^{i\alpha},\qquad
\chi_{\nu^c}\rightarrow \chi_{\nu^c}\,e^{i\beta},\qquad
\overline{\chi}_{\nu^c}\rightarrow \overline{\chi}_{\nu^c}\,e^{i\gamma}
\end{equation}
Then from the requirement that nonzero $a_1^{ij}$ and $b_1$ are allowed
while $a^{i}_2$, $a_3$, $b^i_2$, $c^i_2$  are set to zero by the symmetry:
\begin{eqnarray}
\label{allowed&forbidden}
\alpha+\gamma=\pi k,~~
\beta+\gamma=\frac{2 \pi}{3}l,~~
(\alpha+\gamma)+(\beta+\gamma)\neq 2\pi m,\\
\beta+\gamma\neq \pi n,~~(\alpha+\gamma)+2(\beta+\gamma)\neq 2\pi q,~~
(\alpha+\gamma)+3(\beta+\gamma)\neq 2\pi p.\nonumber
\end{eqnarray}
Conditions (\ref{allowed&forbidden})  imply that $\frac{2}{3}l$
is not integer and, consequently, that $\alpha-\beta\neq 2\pi j$:
\begin{equation}
\alpha-\beta=(\alpha+\gamma)-(\beta+\gamma)=\pi(k-\frac23 l)\neq 2\pi j\neq 0
\end{equation}
In other words, $\nu^c$  and $\chi_{\nu^c}$ have different
transformation properties under the discrete symmetry. If the last term in
(\ref{superpotential}) which is responsible for both small neutrino masses
via the see-saw mechanism and leptogenesis is allowed by this symmetry, then the
term $\chi_{\nu^c}(L_j H_k^{u})$ is necessarily forbidden just as was
assumed.

Bare mass term $M\chi_{\nu^c}\overline{\chi}_{\nu^c}$ is not invariant
under transformations of the discrete symmetry and therefore is not allowed.
From equations (\ref{allowed&forbidden}) it also follows that
$\alpha+\gamma=\pi k$  with $k$ --- odd, so that bare mass term
$M\nu^c\overline{\chi}_{\nu^c}$ is forbidden as well. Finally, coefficients
$b_3$ and $b_4$ of the last two terms in (\ref{NRsuppot}) vanish for the same
reasons.

After $\tilde{\chi}_{\nu^c}$ and $\tilde{\overline{\chi}}_{\nu^c}$  develop
nonzero VEVs, the $U(1)$ symmetry, as well as the discrete symmetry
(\ref{transformations}),
is broken down. Components of chiral (super)fields $\chi_{\nu^c}$,
$\overline{\chi}_{\nu^c}$ and gauge (super)field become massive.
As $m^2_{\chi}=m^2_{\overline{\chi}}$ is assumed, VEVs of
$\tilde{\chi}_{\nu^c}$ and $\tilde{\overline{\chi}}_{\nu^c}$ are equal
and it is natural to  introduce new fields
$h_1=(\chi-\bar{\chi})/\sqrt{2}$ and $h_2=(\chi+\bar{\chi})/\sqrt{2}$.
The imaginary component of $\tilde{h}_1$ is
`eaten up' by the vector gauge field $A$, which acquires large mass
$M_A=M=g q v_0\sim 10^{14}$ GeV. The real component $\eta$ of $\tilde{h}_1$ acquire the
same mass $M_\eta=M$.

From the analysis of gauge interactions alone, it follows that two-component spinors
 $\lambda$ (superpartner of $A$) and $h_1$
(superpartner of $\tilde{h}_1$) form a four--component Dirac spinor
$M_f (h_1\lambda+\lambda^{+}h_1^{+})$ with mass $M_f=M$. Non--renormalizable
interactions induce Majorana--type mass term $\sim m_{soft}(h_1h_1+c.c.)$.
There is
also Majorana--type mass term  $\sim m_{soft}(\lambda\lambda+c.c.)$
coming from the soft
supersymmetry breaking, so that two linear combinations of $h_1$ and
$\lambda$ are Majorana fermions with large ($\sim M$) and close masses.
Non--renormalizable interactions (\ref{NRsuppot}) also induce masses
$\sim m_{soft}$ for the real component of $\tilde{h}_2$ and its fermionic
superpartner $h_2$.

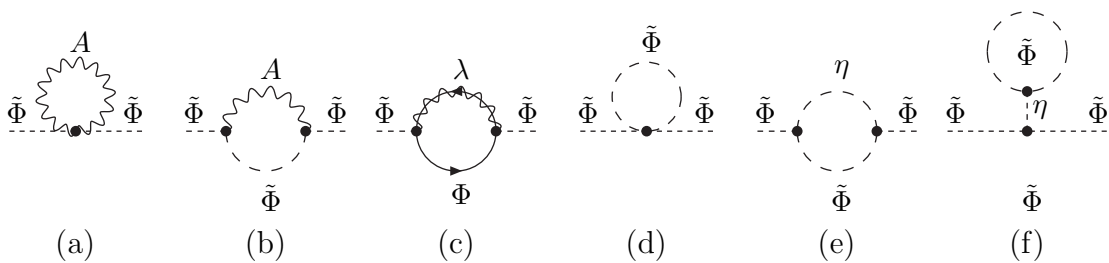
\begin{figure}[h]
\caption{\label{oneloopselfen}
One--loop contributions to self--energy of scalars.}
\begin{center}
\begin{tabular}{cccccc}
~ &~ &~ & ~ & ~ & ~ \\
\begin{picture}(60,60)(0,-10)
\Text(8,23)[bc]{\small $\tilde{\Phi}$} 
\Text(52,23)[bc]{\small $\tilde{\Phi}$} 
\Text(32,50)[bc]{\small $A$} 
\Vertex(30,20){2}
\DashLine(5,20)(30,20){2}
\DashLine(30,20)(55,20){2}
\PhotonArc(30,33)(13,0,360){2}{14} 
\end{picture}
&
\begin{picture}(60,60)(0,-10)
\Text(3,23)[bc]{\small $\tilde{\Phi}$} 
\Text(57,23)[bc]{\small $\tilde{\Phi}$} 
\Text(32,0)[tc]{\small $\tilde{\Phi}$} 
\Text(32,40)[bc]{\small $A$} 
\DashLine(0,20)(15,20){2} 
\DashLine(45,20)(60,20){2} 
\Vertex(15,20){2} \Vertex(45,20){2}
\PhotonArc(30,20)(15,0,180){2}{7} 
\DashCArc(30,20)(15,180,0){4} 
\end{picture}
&
\begin{picture}(60,60)(0,-10)
\Text(3,23)[bc]{\small $\tilde{\Phi}$} 
\Text(57,23)[bc]{\small $\tilde{\Phi}$} 
\Text(32,0)[tc]{\small $\Phi$} 
\Text(32,40)[bc]{\small $\lambda$} 
\DashLine(0,20)(15,20){2} 
\DashLine(45,20)(60,20){2} 
\Vertex(15,20){2} \Vertex(45,20){2}
\ArrowArc(30,20)(15,0,180) 
\PhotonArc(30,20)(15,0,180){2}{7} 
\ArrowArc(30,20)(15,180,0) 
\end{picture}
&
\begin{picture}(60,60)(0,-10)
\Text(8,23)[bc]{\small $\tilde{\Phi}$} 
\Text(52,23)[bc]{\small $\tilde{\Phi}$} 
\Text(32,50)[bc]{\small $\tilde{\Phi}$} 
\Vertex(30,20){2}
\DashLine(5,20)(30,20){2}
\DashLine(30,20)(55,20){2}
\DashCArc(30,33)(13,0,360){4} 
\end{picture}
&
\begin{picture}(60,60)(0,-10)
\Text(3,23)[bc]{\small $\tilde{\Phi}$} 
\Text(57,23)[bc]{\small $\tilde{\Phi}$} 
\Text(32,0)[t]{\small $\tilde{\Phi}$} 
\Text(32,40)[b]{\small $\eta$} 
\Vertex(15,20){2} \Vertex(45,20){2}
\DashLine(0,20)(15,20){2} 
\DashLine(45,20)(60,20){2} 
\DashCArc(30,20)(15,0,180){4} 
\DashCArc(30,20)(15,180,0){4} 
\end{picture}
&
\begin{picture}(60,60)(0,-10)
\Text(3,23)[bc]{\small $\tilde{\Phi}$} 
\Text(57,23)[bc]{\small $\tilde{\Phi}$} 
\Text(32,0)[t]{\small $\tilde{\Phi}$} 
\Text(30,45)[bc]{\small $\tilde{\Phi}$} 
\Text(35,25)[b]{\small $\eta$} 
\Vertex(30,20){2} \Vertex(30,35){2}
\DashLine(0,20)(60,20){2} 
\DashLine(30,20)(30,35){2} 
\DashCArc(30,50)(15,0,180){4} 
\DashCArc(30,50)(15,180,0){4} 
\end{picture}\\
(a) & (b) & (c) & (d) & (e) & (f) \\
\end{tabular}
\end{center}
\end{figure}

The gauge boson $A$, the fermion $\lambda$ and the scalar $\eta$ interact with other
states in the fundamental representation of $E_6$  (in particular with Higgses)
so that the self--energy of scalars receives large  contributions from diagrams with
exchange of these heavy fields. For instance,
see the case of one loop in \fref{oneloopselfen}.

As has already been mentioned in the introduction, the supersymmetric structure of the
Lagrangian ensures that all large corrections, associated with the exchange of heavy fields,
cancel out and only terms proportional to the soft supersymmetry breaking parameters remain:
\begin{equation}
\fl\Pi^{0}(p^2)=4\,g_{\omega^{'}_0}^2\,q_{\tilde{\Phi}}^2\, m^2_{\tilde{\Phi}}
B(m^2_{\tilde{\Phi}}, M^2),\qquad
B(x,y)=\frac{(2\pi\mu)^{2\epsilon}}{\pi^2}\int
\frac{d^{4-2\epsilon} k}{[k^2-x][(k-p)^2-y]}
\end{equation}

\paragraph{$\bf SU_C(3)\times SU_L(2)\times SU_R(2)\times U^2(1)$ models.}
Tables \ref{masslessadj} and \ref{masslessfund}  show, that it
is sufficient to consider only the case of the first charge
assignment and three $SU(3)_R\rightarrow SU_R(2)$ projections, the
other cases being completely analogous.

In the case of the left--right symmetric ($R$) model, the zero modes in
$\delta\bf (27+\overline{27})$ are $SU_R(2)$ doublets
$\chi_{R}=(\chi_{e^c}$,$\chi_{\nu^c}$) and
$\overline{\chi}_R=(\overline{\chi}_{\nu^c}$,$\overline{\chi}_{e^c}$).
The scalar potential coming from gauge interactions (\ref{gaugeint})
and soft supersymmetry breaking is of the form
\begin{eqnarray}
\fl V=\frac18\left(g^2_{Y_L}q^2_{Y_L}+g^2_{Y_R}q^2_{Y_R}\right)
\left[\left(\tilde{\chi}_R^{*}\tilde{\chi}_R\right)-
\left(\tilde{\overline{\chi}}_R^{*}\tilde{\overline{\chi}}_R\right)\right]^2+
\frac{g^2_{W_R}}2\left[\left(\tilde{\chi}_R^{*}\frac{\tau_i}2\tilde{\chi}_R\right)+
\left(\tilde{\overline{\chi}}_R^{*}\frac{\tau_i}2\tilde{\overline{\chi}}_R\right)\right]^2\\
-\left[m^2_{\chi}\left(\tilde{\chi}_R^{*}\tilde{\chi}_R\right)
+m^2_{\overline{\chi}}\left(\tilde{\overline{\chi}}_R^{*}
\tilde{\overline{\chi}}_R\right)\right]\nonumber
\end{eqnarray}
Let $v_1\equiv \langle \tilde{\chi}_{\nu^c} \rangle$, $v_2\equiv
\langle \tilde{\chi}_{e^c} \rangle$ and $u_1\equiv \langle
\tilde{\overline{\chi}}_{\nu^c} \rangle$, $u_2\equiv \langle
\tilde{\overline{\chi}}_{e^c} \rangle$. The corresponding classical
potential is
\begin{eqnarray}
\label{LRclasspot} \fl
V=\frac18\left(g^2_{Y_L}q^2_{Y_L}+g^2_{Y_R}q^2_{Y_R}\right)
\left[\left(|v_1|^2+|v_2|^2\right)-\left(|u_1|^2+|u_2|^2\right)\right]\\
+ \frac12\, g^2_{W_R}|\left[v_2^{*}v_1+u_1^{*}u_2\right]|^2
+\frac18\, g^2_{W_R}\left[\left(|v_2|^2-|v_1|^2\right)+
\left(|u_1|^2-|u_2|^2\right)\right]\nonumber\\-
\left[m^2_{\chi}\left(|v_1|^2+|v_2|^2\right)+
m^2_{\overline{\chi}}\left(|u_1|^2+|u_2|^2\right)\right]\nonumber
\end{eqnarray}

There are symmetry breaking directions, which are $D$ -- flat. Vanishing of
the first and the third term in (\ref{LRclasspot}) requires that
$|v_1|=|u_1|$ and $|v_2|=|u_2|$. With these conditions satisfied, the second
term in (\ref{LRclasspot}) vanishes if  $v_1=0$ or $v_2=0$ or $arg(u_2)+arg(v_2)-
arg(u_1)-arg(v_1)=\pi$.

The non-renormalizable superpotential, which bounds the classical potential from below,
is similar to (\ref{NRsuppot}) with $\nu^c$,
$\chi_{\nu^c}$ and $\overline{\chi}_{\nu^c}$ replaced with $SU_R(2)$
doublets $\psi_R=(e^c,\nu^c)$, $\chi_R=(\chi_{e^c},\chi_{\nu^c})$
and
$\overline{\chi}_R=(\overline{\chi}_{\nu^c},\overline{\chi}_{e^c})$.
Its explicit form is

\begin{eqnarray}
\fl
W=M_c^{-1}\left[a_1^{ij}(\psi_R^i\overline{\chi}_R)(\psi_R^j\overline{\chi}_R)+
a_2^i(\psi_R^i\overline{\chi}_R)(\psi_R^i\overline{\chi}_R)+
a_3(\chi_R\overline{\chi}_R)(\chi_R\overline{\chi}_R)\right]\\
+M_c^{-3}\left[(\chi_R\overline{\chi}_R)(\chi_R\overline{\chi}_R)
(\chi_R\overline{\chi}_R)+b^i (\psi_R^i\overline{\chi}_R)
(\chi_R\overline{\chi}_R)(\chi_R\overline{\chi}_R)\right.\nonumber\\
\left. +b_3^{ij}(\psi_R^i\overline{\chi}_R)(h^j\overline{\chi}_R)
(\chi_R\overline{\chi}_R)+b_4^{ijk}(\psi_R^i\overline{\chi}_R)
(\psi_R^j\overline{\chi}_R)(\psi_R^k\overline{\chi}_R)\right]+\dots
\nonumber
\end{eqnarray}

The freedom of $SU_R(2)$ gauge transformations allows to rotate away
a possible VEV for one of the isospin components of one of the
scalar fields, so one can take $u_2=0$ at the minimum of the
potential.  Since the classical potential under consideration
reaches its minimum in one of the $D$ -- flat directions, $v_2$ is
equal to zero as well. Then the following analysis is the same as in
the case of the $SU_C(3)\times SU_L(2)\times U^3(1)$ model and furnishes
the same result. The symmetry is broken down to $SU_C(3)\times
SU_L(2)\times U^2(1)$.

The second skew left--right symmetric ($R^{''}$) model differs from
the one above in $S$ instead of $e^c$ being an $SU_{R^{''}}(2)$
counterpart of the right--handed neutrino: $\chi_{R^{''}}=(\chi_{\nu^c},
\chi_S$) and $\overline{\chi}_{R^{''}}=(\overline{\chi}_{S},
\overline{\chi}_{\nu^c}$). The symmetry is broken down to
$SU_C(3)\times SU_L(2)\times U^2(1)$ as well.

In the case of the first skew left--right symmetric ($R^{'}$) model, the
right--handed neutrino is  an $SU_{R^{'}}(2)$ singlet. Massless states
in $\delta (27+\overline{27})$ are $\chi_{\nu^c}$,
$\chi_{R^{'}}=(\chi_{H^u}$,$\chi_{L}$) and
$\overline{\chi}_{\nu^c}$,
$\overline{\chi}_{R^{'}}=(\overline{\chi}_{L}$,$\overline{\chi}_{H^u}$).
The classical potential coming from renormalizable and nonrenormalizable
interactions is similar to (\ref{classicalpot})
and yields the same results. The symmetry is broken down to
$SU_C(3)\times SU_L(2)\times SU_{R^{'}}(2)\times U(1)$.

Since the Higgs doublets $(\chi_{H^u}$,$\chi_{L}$) and
$(\overline{\chi}_{L}$,$\overline{\chi}_{H^u}$) are contained in
$\delta\bf(27+\overline{27})$ as zero modes, there are directions in which the $D$-term potential vanishes for whatever large VEVs
of these fields, i.e., there is a risk of breaking electroweak symmetry at a very high scale. To
avoid it, the part of the classical potential which comes from the soft supersymmetry breaking
should be positive in those directions, as is the case in the MSSM.

The coexistence of all the terms in the second row of
(\ref{superpotential}) leads to the rapid proton decay,
mediated by new
$D$ and $D^c$ quarks, unless those are very heavy. VEV of $S$ which
gives masses $\lambda_5^{ijk}\langle S_i\rangle$ to $D$ and $D^c$ is
also the source of $\mu$--term $\lambda_4^{ijk}\langle S_i\rangle$.
Although neither for $G_{U(1)}$ nor for $G_{SU_R(2)}$ the couplings
$\lambda_5^{ijk}$ and $\lambda_5^{ijk}$ are related by symmetry, it is
not natural to expect $D$ and $D^c$ to be much heavier than 1 TeV
which is insufficient to suppress the proton decay. A solution to
this problem may be provided by an appropriate  discrete symmetry,
which forbids some of the couplings in (\ref{superpotential}).

\begin{table}[h]
\begin{flushright}
\caption{\label{discrsymm}Transformation properties of components of
$SU_R(2)$ doublets.} \footnotesize
\begin{tabular}{lllllllllllllllllllll}
\br Model & \centre{6}{$R$} & \centre{6}{$R^{'}$} &
\centre{6}{$R^{''}$}\\
\ns\ns&\crule{2}&\crule{2}&\crule{2}&\crule{2}&\crule{2}&\crule{2}
&\crule{2}&\crule{2}&\crule{2}\\
& $d^c$ & $u^c$ & $H^u$ & $H^d$ & $e^c$ & $\nu^c$  & $D^c$ & $u^c$ &
$H^u$ & $L$ & $e^c$ & $S$ & $D^c$ & $d^c$ & $H^d$ & $L$ & $\nu^c$ & $S$\\
1 & + & + & + & + & - & - & - & + & + & - & - & + & - & + & + & - & - & + \\
2 & + & + & + & + & - & - & + & + & + & - & - & + & + & + & + & - & - & +\\
\br
\end{tabular}
\end{flushright}
\end{table}

If this is a $Z_2$ symmetry, then there are only two models
\cite{Hambye:2000bn,Ma:1987jg} compatible with leptogenesis and
nonzero neutrino masses. In the first model $L$ and $e^c$, $\nu^c$,
$D$, $D^c$ are odd, while the rest of the states are even, so that
$\lambda_9=\lambda_{10}=0$. The second model differs from the first
in $D$ and $D^c$ being even, so that
$\lambda_6=\lambda_7=\lambda_8=0$. Transformations of the discrete
symmetry should commute with transformations of the gauge symmetry.
\Tref{discrsymm} shows, that this condition is satisfied only in the
case of the left--right symmetric ($R$) model, while in both skew
left--right symmetric models components of $SU_R(2)$ doublets
transform differently. Therefore, if proton stability is assured by
such a discrete symmetry, the only allowed gauge group after breaking
of $G$ is $G^{'}_{U(1)}=SU_C(3)\times SU_L(2)\times U^2(1)$.

\section{\label{conclusions}Conclusions}
In this paper, an attempt to derive `low--energy' extensions of the SM,
compatible with the Fukugita--Yanagida scenario, from a superstring inspired
$E_6$ model has been made.

This model allows for six charge assignments compatible with the standard
model. Charge conservation in the processes involving states of different
generations requires that the same charge assignment be used for all
generations.

As is discussed above, the gauge symmetry is broken in a sequence of stages. The first
stage is due to Calabi--Yau compactification and the effective Higgs mechanism. The condition
that the proton is long-lived requires that the symmetry is broken either to $G_{U(1)}$ or $G_{SU_R(2)}$.

As the temperature goes down, supersymmetry breaks down spontaneously in the hidden
sector. Supersymmetry breaking is mediated to the visible sector by gravity and manifests
itself in the soft terms.

At the next stage right--handed scalar neutrinos and right--handed neutrinos of two
additional generations
$\bf (27+\overline{27})$ develop nonzero VEV, breaking the $B-L$ symmetry.
The introduced simple discrete symmetry ensures that $B-L$ is broken at the scale which
is sufficiently high for generating large masses of the right--handed neutrinos and that right--handed
scalar neutrinos of the three known generations do not acquire VEV. It also forbids
Yukawa couplings which, if present, would induce large masses of the conventional neutrinos.
The supersymmetric structure of the theory ensures that large quantum corrections to masses
of scalars, associated with the presence of heavy gauge fields, cancel out. Provided that a
rapid proton decay mediated by the new quarks is forbidden by a $Z_2$ symmetry, the sought `low--energy' gauge group is
given by $G^{'}_{U(1)}$ for both allowed intermediate scale gauge groups
$G_{U(1)}$ and $G_{SU_R(2)}$.

Apart from the additional (compared to the SM) $U(1)$ symmetry, the characteristic
feature of this `low--energy' model is the extended particle content. In addition to the known
particles and right--handed neutrinos it contains the SM singlets S, new heavy quarks and three
generations of Higgses, as well as their superpartners. Since both new quarks and Higgses
couple to right--handed neutrinos, there are more $B-L$ violating decay channels than in the SM
or its supersymmetric extension. At the same time there are more processes which wash the
generated $B-L$ asymmetry out. The interplay of these effects requires further investigation.

\ack
I would like to thank Professor E A Paschos for inspiration and useful discussions. Financial
support from the Graduiertenkolleg 841 Physik der Elementarteilchen an Beschleunigern und
im Universum at Dortmund University is gratefully acknowledged.

\begin{table}[h]
\begin{flushright}
\caption{\label{weights_of_27}Weights  of {\bf 27} of $E_6$. Here
$\varepsilon=\frac13(\alpha+\beta+\gamma)$, $\bar{x}\equiv-x$. The zero root
breaking SM gauge group preserving direction is given by
$Z=(-\varepsilon, \varepsilon, \beta, 2\varepsilon-\gamma, \varepsilon, 0)$.}
\footnotesize
\begin{tabular}{lllllllllllllll}
\br
 $E_6$ & \centre{3}{$SU(3)$} & \centre{4}{$SU(2)_{,Y}$} & \centre{6}{Assignments} & $(Z,\rho)$ \\
\ns\ns & \crule{3}& \crule{4}& \crule{6}& \\
  & $C$ & $L$ & $R$ & $L$ & $R$ & $R^{'}$ & $R^{''}$ & I & II & III & IV & V & VI & \\
\ns\mr
{\bf 27} & \textbf{3} & \textbf{3} & \textbf{1} & \textbf{2} & \centre{3}{~}  & \centre{6}{~} & \\
\ns\mr
 [$100000$] & [$10$] & [$10$] & [$00$] & $[1],_{1}$ & $[0],_{0}$ & $[0],_{0}$ & $[0],_{0}$ & u & u & u & u & u & u & $-\varepsilon$\\

 [$1\bar{1}0010$] & [$\bar{1}1$] & [$10$] & [$00$] & $[1],_{1}$ & $[0],_{0}$ & $[0],_{0}$ & $[0],_{0}$ & u & u & u & u & u & u & $-\varepsilon$\\

 [$10000\bar{1}$] & [$0\bar{1}$] & [$10$] & [$00$] & $[1],_{1}$ & $[0],_{0}$ & $[0],_{0}$ & $[0],_{0}$ & u & u & u & u & u & u & $-\varepsilon$\\

 [$0000\bar{1}1$] & [$10$] & [$\bar{1}1$] & [$00$] & $[\bar{1}],_{1}$ & $[0],_{0}$ & $[0],_{0}$ & $[0],_{0}$ & d & d & d & d & d & d & $-\varepsilon$\\

 [$0\bar{1}0001$] & [$\bar{1}1$] & [$\bar{1}1$] & [$00$] & $[\bar{1}],_{1}$ & $[0],_{0}$ & $[0],_{0}$ & $[0],_{0}$ & d & d & d & d & d & d & $-\varepsilon$\\

 [$0000\bar{1}0$] & [$0\bar{1}$] & [$\bar{1}1$] & [$00$] & $[\bar{1}],_{1}$ & $[0],_{0}$ & $[0],_{0}$ & $[0],_{0}$ & d & d & d & d & d & d & $-\varepsilon$\\
\mr
{\bf 27}  & \textbf{3} & \textbf{3} & \textbf{1} & \textbf{1} & \centre{3}{~} & \centre{6}{~} & \\
\ns\mr
[$\bar{1}10000$] & [$10$] & [$0\bar{1}$] & [$00$] & $[0],_{\bar{2}}$ & $[0],_{0}$ & $[0],_{0}$ & $[0],_{0}$ & D & D & D & D & D & D & $2\varepsilon$\\

 [$\bar{1}00010$] & [$\bar{1}1$] & [$0\bar{1}$] & [$00$] & $[0],_{\bar{2}}$ & $[0],_{0}$ & $[0],_{0}$ & $[0],_{0}$ & D & D & D & D & D & D & $2\varepsilon$\\

 [$\bar{1}1000\bar{1}$] & [$0\bar{1}$] & [$0\bar{1}$] & [$00$] & $[0],_{\bar{2}}$ & $[0],_{0}$ & $[0],_{0}$ & $[0],_{0}$ & D & D & D & D & D & D & $2\varepsilon$\\
 \mr
{\bf 27}  & $\bf\bar{3}$ & \textbf{1} & $\bf\bar{3}$ &  \textbf{1} & \centre{2}{~} & &\centre{5}{~} & & \\
\ns\mr  [$00\bar{1}101$] & [$01$] & [$00$] & [$\bar{1}0$] & $[0],_{0}$ & $[\bar{1}],_{\bar{1}}$ & $[\bar{1}],_{\bar{1}}$ & $[0],_{2}$ & $u^c$ & $u^c$ & $d^c$ & $D^c$ & $d^c$ & $D^c$ & $-\varepsilon+\alpha$\\

 [$01\bar{1}1\bar{1}0$] & [$1\bar{1}$] & [$00$] & [$\bar{1}0$] & $[0],_{0}$ & $[\bar{1}],_{\bar{1}}$ & $[\bar{1}],_{\bar{1}}$ & $[0],_{2}$ & $u^c$ & $u^c$ & $d^c$ & $D^c$ & $d^c$ & $D^c$ & $-\varepsilon+\alpha$\\

 [$00\bar{1}100$] & [$\bar{1}0$] & [$00$] & [$\bar{1}0$] & $[0],_{0}$ & $[\bar{1}],_{\bar{1}}$ & $[\bar{1}],_{\bar{1}}$ & $[0],_{2}$ & $u^c$ & $u^c$ & $d^c$ & $D^c$ & $d^c$ & $D^c$ & $-\varepsilon+\alpha$\\
\mr
 [$000\bar{1}11$] & [$01$] & [$00$] & [$01$] & $[0],_{0}$ & $[0],_{2}$ & $[1],_{\bar{1}}$ & $[1],_{\bar{1}}$ & $D^c$ & $d^c$ & $D^c$ & $d^c$ & $u^c$ & $u^c$ & $-\varepsilon+\gamma$\\

 [$010\bar{1}00$] & [$1\bar{1}$] & [$00$] & [$01$] & $[0],_{0}$ & $[0],_{2}$ & $[1],_{\bar{1}}$ & $[1],_{\bar{1}}$ & $D^c$ & $d^c$ & $D^c$ & $d^c$ & $u^c$ & $u^c$ & $-\varepsilon+\gamma$\\

 [$000\bar{1}10$] & [$\bar{1}0$] & [$00$] & [$01$] & $[0],_{0}$ & $[0],_{2}$ & $[1],_{\bar{1}}$ & $[1],_{\bar{1}}$ & $D^c$ & $d^c$ & $D^c$ & $d^c$ & $u^c$ & $u^c$ & $-\varepsilon+\gamma$\\
\mr
 [$0\bar{1}1000$] & [$01$] & [$00$] & [$1\bar{1}$] & $[0],_{0}$ & $[1],_{\bar{1}}$ & $[0],_{2}$ & $[\bar{1}],_{\bar{1}}$ & $d^c$ & $D^c$ & $u^c$ & $u^c$ & $D^c$ & $d^c$ & $-\varepsilon+\beta$\\

 [$0010\bar{1}\bar{1}$] & [$1\bar{1}$] & [$00$] & [$1\bar{1}$] & $[0],_{0}$ & $[1],_{\bar{1}}$ & $[0],_{2}$ & $[\bar{1}],_{\bar{1}}$ & $d^c$ & $D^c$ & $u^c$ & $u^c$ & $D^c$ & $d^c$ & $-\varepsilon+\beta$\\

 [$0\bar{1}100\bar{1}$] & [$\bar{1}0$] & [$00$] & [$1\bar{1}$] & $[0],_{0}$ & $[1],_{\bar{1}}$ & $[0],_{2}$ & $[\bar{1}],_{\bar{1}}$ & $d^c$ & $D^c$ & $u^c$ & $u^c$ & $D^c$ & $d^c$ & $-\varepsilon+\beta$\\
 \mr
{\bf 27}  &  \textbf{1} & $\bf\bar{3}$ & \textbf{3} & \textbf{2} & \centre{2}{~} & & \centre{5}{~} & &  \\
\ns\mr
[$001\bar{1}1\bar{1}$] & [$00$] & [$1\bar{1}$] & [$10$] & $[1],_{\bar{1}}$ & $[1],_{1}$ & $[1],_{1}$ & $[0],_{\bar{2}}$ & $E^c$ & $E^c$ & N & $\nu$ & N & $\nu$ & $2\varepsilon-\alpha$\\

 [$\bar{1}01\bar{1}00$] & [$00$] & [$\bar{1}0$] & [$10$] & $[\bar{1}],_{\bar{1}}$ & $[1],_{1}$ & $[1],_{1}$ & $[0],_{\bar{2}}$ & $N^c$ & $N^c$ & E & e & E & e & $2\varepsilon-\alpha$\\
\mr
 [$01\bar{1}010$] & [$00$] & [$1\bar{1}$] & [$\bar{1}1$] & $[1],_{\bar{1}}$ & $[\bar{1}],_{1}$ & $[0],_{\bar{2}}$ & $[1],_{1}$ & N & $\nu$ & $E^c$ & $E^c$ & $\nu$ & N & $2\varepsilon-\beta$\\

 [$\bar{1}1\bar{1}001$] & [$00$] & [$\bar{1}0$] & [$\bar{1}1$] & $[\bar{1}],_{\bar{1}}$ & $[\bar{1}],_{1}$ & $[0],_{\bar{2}}$ & $[1],_{1}$ & E & e & $N^c$ & $N^c$ & e & E & $2\varepsilon-\beta$\\
\mr
 [$00010\bar{1}$] & [$00$] & [$1\bar{1}$] & [$0\bar{1}$] & $[1],_{\bar{1}}$ & $[0],_{\bar{2}}$ & $[\bar{1}],_{1}$ & $[\bar{1}],_{1}$ & $\nu$ & N & $\nu$ & N & $E^c$ & $E^c$ & $2 \varepsilon-\gamma$\\

 [$\bar{1}001\bar{1}0$] & [$00$] & [$\bar{1}0$] & [$0\bar{1}$] & $[\bar{1}],_{\bar{1}}$ & $[0],_{\bar{2}}$ & $[\bar{1}],_{1}$ & $[\bar{1}],_{1}$ & e & E & e & E & $N^c$ & $N^c$ & $2 \varepsilon-\gamma$\\
\mr
{\bf 27}  & \textbf{1} & $\bf\bar{3}$ & \textbf{3} & \textbf{1} & \centre{2}{~} & & \centre{5}{~} & &  \\
\ns\mr
[$1\bar{1}1\bar{1}00$] & [$00$] & [$01$] & [$10$] & $[0],_{2}$ & $[1],_{1}$ & $[1],_{1}$ & $[0],_{\bar{2}}$ & $e^c$ & $e^c$ & $\nu^c$ & S & $\nu^c$ & S & $-\varepsilon-\alpha$\\
\mr
 [$10\bar{1}001$] & [$00$] & [$01$] & [$\bar{1}1$] & $[0],_{2}$ & $[\bar{1}],_{1}$ & $[0],_{\bar{2}}$ & $[1],_{1}$ & $\nu^c$ & S & $e^c$ & $e^c$ & S & $\nu^c$ & $-\varepsilon-\beta$\\
\mr
 [$1\bar{1}01\bar{1}0$] & [$00$] & [$01$] & [$0\bar{1}$] & $[0],_{2}$ & $[0],_{\bar{2}}$ & $[\bar{1}],_{1}$ & $[\bar{1}],_{1}$ & S & $\nu^c$ & S & $\nu^c$ & $e^c$ & $e^c$ & $-\varepsilon-\gamma$\\
\br
\end{tabular}
\end{flushright}
\end{table}

\begin{table}
\begin{flushright}
\caption{\label{weights_of_78}Weights of {\bf 78} of $E_6$ (beginning). Color, isospin
and hypercharge (as well as $B-L$, unless otherwise indicated) of
the states designated like the SM states are the same as charges of
their SM counterparts. If present, subscript is trice the $B-L$
charge. All charges of overlined states are opposite to those of non
overlined. For example, for $\overline{d_{\bar{2}}}$: $I_3=\frac12$,
$Y=-\frac13$, $B-L=\frac23$.}
\footnotesize
\begin{tabular}{lllllllllllllll}
\br
$E_6$ & \centre{3}{$SU(3)$} & \centre{4}{$SU(2)_{,Y}$} & \centre{6}{Assignments} & $(Z,\rho)$ \\
\ns\ns  & \crule{3} & \crule{4} & \crule{6}\\
  & $C$ & $L$ & $R$ & $L$ & $R$ & $R^{'}$ & $R^{''}$ & I & II & III & IV & V & VI & \\
\ns\mr
{\bf 78} & {\bf 1} & {\bf 8} & {\bf 1}\\
\ns\mr
 [$2\bar{1}0000$] & [$00$] & [$11$] & [$00$] & [$1$],$_{3}$ & [$0$],$_{0}$ & [$0$],$_{0}$ & [$0$],$_{0}$ & $\overline{e}$ & $\overline{e}$ & $\overline{e}$ & $\overline{e}$ & $\overline{e}$ & $\overline{e}$ & $-3\varepsilon$\\

 [$1\bar{1}00\bar{1}1$] & [$00$] & [$\bar{1}2$] & [$00$] & [$\bar{1}$],$_{3}$ & [$0$],$_{0}$ & [$0$],$_{0}$ & [$0$],$_{0}$ & $\overline{\nu}$ & $\overline{\nu}$ & $\overline{\nu}$ & $\overline{\nu}$ & $\overline{\nu}$ & $\overline{\nu}$ & $-3\varepsilon$\\
\mr
 [$10001\bar{1}$] & [$00$] & [$2\bar{1}$] & [$00$] & [$2$],$_{0}$ & [$0$],$_{0}$ & [$0$],$_{0}$ & [$0$],$_{0}$ & \scriptsize{$W^+$} & \scriptsize{$W^+$} & \scriptsize{$W^+$} & \scriptsize{$W^+$} & \scriptsize{$W^+$} & \scriptsize{$W^+$} & $0$\\

 [$000000$] & [$00$] & [$00$] & [$00$] & [$0$],$_{0}$ & [$0$],$_{0}$ & [$0$],$_{0}$ & [$0$],$_{0}$ & \scriptsize{$W^0$} & \scriptsize{$W^0$} & \scriptsize{$W^0$} & \scriptsize{$W^0$} & \scriptsize{$W^0$} & \scriptsize{$W^0$} & $0$\\

 [$\bar{1}000\bar{1}1$] & [$00$] & [$\bar{2}1$] & [$00$] & [$\bar{2}$],$_{0}$ & [$0$],$_{0}$ & [$0$],$_{0}$ & [$0$],$_{0}$ & \scriptsize{$W^-$} & \scriptsize{$W^-$} & \scriptsize{$W^-$} & \scriptsize{$W^-$} & \scriptsize{$W^-$} & \scriptsize{$W^-$} & $0$\\
\mr
 [$\bar{1}1001\bar{1}$] & [$00$] & [$1\bar{2}$] & [$00$] & [$1$],$_{\bar{3}}$ & [$0$],$_{0}$ & [$0$],$_{0}$ & [$0$],$_{0}$ & $\nu$ & $\nu$ & $\nu$ & $\nu$ & $\nu$ & $\nu$ & $3\varepsilon$\\

 [$\bar{2}10000$] & [$00$] & [$\bar{1}\bar{1}$] & [$00$] & [$\bar{1}$],$_{\bar{3}}$ & [$0$],$_{0}$ & [$0$],$_{0}$ & [$0$],$_{0}$ & $e$ & $e$ & $e$ & $e$ & $e$ & $e$ & $3\varepsilon$\\
\mr
 [$000000$] & [$00$] & [$00$] & [$00$] & [$0$],$_{0}$ & [$0$],$_{0}$ & [$0$],$_{0}$ & [$0$],$_{0}$ & $\gamma^0$ & $\gamma^0$ & $\gamma^0$ & $\gamma^0$ & $\gamma^0$ & $\gamma^0$ & $0$\\
\mr
{\bf 78} & {\bf 1} & {\bf 1} & {\bf 8} \\
\ns \mr [$001\bar{2}10$] & [$00$] & [$00$] & [$11$] & [$0$],$_{0}$ & [$1$],$_{3}$ & [$2$],$_{0}$ & [$1$],$_{\bar{3}}$ & $e^c$ & $e^c_0$ & $\nu^c$ & $\overline{\nu^c}$ & $\overline{e^c_0}$ & $\overline{e^c}$ & $\gamma-\alpha$\\

 [$01\bar{1}\bar{1}11$] & [$00$] & [$00$] & [$\bar{1}2$] & [$0$],$_{0}$ & [$\bar{1}$],$_{3}$ & [$1$],$_{\bar{3}}$ & [$2$],$_{0}$ & $\nu^c$ & $\overline{\nu^c}$ & $e^c$ & $e^c_0$ & $\overline{e^c}$ & $\overline{e^c_0}$ & $\gamma-\beta$\\

 [$0\bar{1}2\bar{1}0\bar{1}$] & [$00$] & [$00$] & [$2\bar{1}$] & [$0$],$_{0}$ & [$2$],$_{0}$ & [$1$],$_{3}$ & [$\bar{1}$],$_{\bar{3}}$ & $e^c_0$ & $e^c$ & $\overline{e^c_0}$ & $\overline{e^c}$ & $\nu^c$ & $\overline{\nu^c}$ & $\beta-\alpha$\\

 [$000000$] & [$00$] & [$00$] & [$00$] & [$0$],$_{0}$ & [$0$],$_{0}$ & [$0$],$_{0}$ & [$0$],$_{0}$ & $\phi^0$ & $\phi^0$ & $\phi^0$ & $\phi^0$ & $\phi^0$ & $\phi^0$ & $0$\\

 [$01\bar{2}101$] & [$00$] & [$00$] & [$\bar{2}1$] & [$0$],$_{0}$ & [$\bar{2}$],$_{0}$ & [$\bar{1}$],$_{\bar{3}}$ & [$1$],$_{3}$ & $\overline{e^c_0}$ & $\overline{e^c}$ & $e^c_0$ & $e^c$ & $\overline{\nu^c}$ & $\nu^c$ & $\alpha-\beta$\\

 [$0\bar{1}11\bar{1}\bar{1}$] & [$00$] & [$00$] & [$1\bar{2}$] & [$0$],$_{0}$ & [$1$],$_{\bar{3}}$ & [$\bar{1}$],$_{3}$ & [$\bar{2}$],$_{0}$ & $\overline{\nu^c}$ & $\nu^c$ & $\overline{e^c}$ & $\overline{e^c_0}$ & $e^c$ & $e^c_0$ & $\beta-\gamma$\\

 [$00\bar{1}2\bar{1}0$] & [$00$] & [$00$] & [$\bar{1}\bar{1}$] & [$0$],$_{0}$ & [$\bar{1}$],$_{\bar{3}}$ & [$\bar{2}$],$_{0}$ & [$\bar{1}$],$_{3}$ & $\overline{e^c}$ & $\overline{e^c_0}$ & $\overline{\nu^c}$ & $\nu^c$ & $e^c_0$ & $e^c$ & $\alpha-\gamma$\\
\mr
 [$000000$] & [$00$] & [$00$] & [$00$] & [$0$],$_{0}$ & [$0$],$_{0}$ & [$0$],$_{0}$ & [$0$],$_{0}$ & $\omega^0$ & $\omega^0$ & $\omega^0$ & $\omega^0$ & $\omega^0$ & $\omega^0$ & $0$\\
\mr
{\bf 78} & {\bf 8} & {\bf 1} & {\bf 1}\\
\ns\mr
 [$000001$] & [$11$] & [$00$] & [$00$] & [$0$],$_{0}$ & [$0$],$_{0}$ & [$0$],$_{0}$ & [$0$],$_{0}$ & $g$ & $g$ & $g$ & $g$ & $g$ & $g$ & $0$\\

 [$0\bar{1}0011$] & [$\bar{1}2$] & [$00$] & [$00$] & [$0$],$_{0}$ & [$0$],$_{0}$ & [$0$],$_{0}$ & [$0$],$_{0}$ & $g$ & $g$ & $g$ & $g$ & $g$ & $g$ & $0$\\

 [$0100\bar{1}0$] & [$2\bar{1}$] & [$00$] & [$00$] & [$0$],$_{0}$ & [$0$],$_{0}$ & [$0$],$_{0}$ & [$0$],$_{0}$ & $g$ & $g$ & $g$ & $g$ & $g$ & $g$ & $0$\\

 [$000000$] & [$00$] & [$00$] & [$00$] & [$0$],$_{0}$ & [$0$],$_{0}$ & [$0$],$_{0}$ & [$0$],$_{0}$ & $g$ & $g$ & $g$ & $g$ & $g$ & $g$ & $0$\\

 [$000000$] & [$00$] & [$00$] & [$00$] & [$0$],$_{0}$ & [$0$],$_{0}$ & [$0$],$_{0}$ & [$0$],$_{0}$ & $g$ & $g$ & $g$ & $g$ & $g$ & $g$ & $0$\\

 [$0100\bar{1}\bar{1}$] & [$1\bar{2}$] & [$00$] & [$00$] & [$0$],$_{0}$ & [$0$],$_{0}$ & [$0$],$_{0}$ & [$0$],$_{0}$ & $g$ & $g$ & $g$ & $g$ & $g$ & $g$ & $0$\\

 [$0\bar{1}0010$] & [$\bar{2}1$] & [$00$] & [$00$] & [$0$],$_{0}$ & [$0$],$_{0}$ & [$0$],$_{0}$ & [$0$],$_{0}$ & $g$ & $g$ & $g$ & $g$ & $g$ & $g$ & $0$\\

 [$00000\bar{1}$] & [$\bar{1}\bar{1}$] & [$00$] & [$00$] & [$0$],$_{0}$ & [$0$],$_{0}$ & [$0$],$_{0}$ & [$0$],$_{0}$ & $g$ & $g$ & $g$ & $g$ & $g$ & $g$ & $0$\\
\mr
{\bf 78} & $\bf\overline{3}$ & {\bf 3} & {\bf 3} \\
\ns\mr
 [$1\bar{1}1\bar{1}10$] & [$01$] & [$10$] & [$10$] & [$1$],$_{1}$ & [$1$],$_{1}$ & [$1$],$_{1}$ & [$0$],$_{\bar{2}}$ & $\overline{X}$ & $\overline{X}$ & $\overline{d_{\bar{2}}}$ & $\overline{d}$ & $\overline{d_{\bar{2}}}$ & $\overline{d}$ & $-\alpha$\\

 [$101\bar{1}0\bar{1}$] & [$1\bar{1}$] & [$10$] & [$10$] & [$1$],$_{1}$ & [$1$],$_{1}$ & [$1$],$_{1}$ & [$0$],$_{\bar{2}}$ & $\overline{X}$ & $\overline{X}$ & $\overline{d_{\bar{2}}}$ & $\overline{d}$ & $\overline{d_{\bar{2}}}$ & $\overline{d}$ & $-\alpha$\\

 [$1\bar{1}1\bar{1}1\bar{1}$] & [$\bar{1}0$] & [$10$] & [$10$] & [$1$],$_{1}$ & [$1$],$_{1}$ & [$1$],$_{1}$ & [$0$],$_{\bar{2}}$ & $\overline{X}$ & $\overline{X}$ & $\overline{d_{\bar{2}}}$ & $\overline{d}$ & $\overline{d_{\bar{2}}}$ & $\overline{d}$ & $-\alpha$\\

 [$10\bar{1}011$] & [$01$] & [$10$] & [$\bar{1}1$] & [$1$],$_{1}$ & [$\bar{1}$],$_{1}$ & [$0$],$_{\bar{2}}$ & [$1$],$_{1}$ & $\overline{d_{\bar{2}}}$ & $\overline{d}$ & $\overline{X}$ & $\overline{X}$ & $\overline{d}$ & $\overline{d_{\bar{2}}}$ & $-\beta$\\

 [$11\bar{1}000$] & [$1\bar{1}$] & [$10$] & [$\bar{1}1$] & [$1$],$_{1}$ & [$\bar{1}$],$_{1}$ & [$0$],$_{\bar{2}}$ & [$1$],$_{1}$ & $\overline{d_{\bar{2}}}$ & $\overline{d}$ & $\overline{X}$ & $\overline{X}$ & $\overline{d}$ & $\overline{d_{\bar{2}}}$ & $-\beta$\\

 [$10\bar{1}010$] & [$\bar{1}0$] & [$10$] & [$\bar{1}1$] & [$1$],$_{1}$ & [$\bar{1}$],$_{1}$ & [$0$],$_{\bar{2}}$ & [$1$],$_{1}$ & $\overline{d_{\bar{2}}}$ & $\overline{d}$ & $\overline{X}$ & $\overline{X}$ & $\overline{d}$ & $\overline{d_{\bar{2}}}$ & $-\beta$\\

 [$1\bar{1}0100$] & [$01$] & [$10$] & [$0\bar{1}$] & [$1$],$_{1}$ & [$0$],$_{\bar{2}}$ & [$\bar{1}$],$_{1}$ & [$\bar{1}$],$_{1}$ & $\overline{d}$ & $\overline{d_{\bar{2}}}$ & $\overline{d}$ & $\overline{d_{\bar{2}}}$ & $\overline{X}$ & $\overline{X}$ & $-\gamma$\\

 [$1001\bar{1}\bar{1}$] & [$1\bar{1}$] & [$10$] & [$0\bar{1}$] & [$1$],$_{1}$ & [$0$],$_{\bar{2}}$ & [$\bar{1}$],$_{1}$ & [$\bar{1}$],$_{1}$ & $\overline{d}$ & $\overline{d_{\bar{2}}}$ & $\overline{d}$ & $\overline{d_{\bar{2}}}$ & $\overline{X}$ & $\overline{X}$ & $-\gamma$\\

 [$1\bar{1}010\bar{1}$] & [$\bar{1}0$] & [$10$] & [$0\bar{1}$] & [$1$],$_{1}$ & [$0$],$_{\bar{2}}$ & [$\bar{1}$],$_{1}$ & [$\bar{1}$],$_{1}$ & $\overline{d}$ & $\overline{d_{\bar{2}}}$ & $\overline{d}$ & $\overline{d_{\bar{2}}}$ & $\overline{X}$ & $\overline{X}$ & $-\gamma$\\

 [$0\bar{1}1\bar{1}01$] & [$01$] & [$\bar{1}1$] & [$10$] & [$\bar{1}$],$_{1}$ & [$1$],$_{1}$ & [$1$],$_{1}$ & [$0$],$_{\bar{2}}$ & $\overline{Y}$ & $\overline{Y}$ & $\overline{u_{\bar{2}}}$ & $\overline{u}$ & $\overline{u_{\bar{2}}}$ & $\overline{u}$ & $-\alpha$\\
\br
\end{tabular}
\end{flushright}
\end{table}

\addtocounter{table}{-1}

\begin{table}
\begin{flushright}
\caption{Weights of {\bf 78} of $E_6$ (continuation).}
\footnotesize
\begin{tabular}{lllllllllllllll}
\br
{\bf 78} & $\bf\overline{3}$ & {\bf 3} & {\bf 3}\\
\ns \mr
 [$001\bar{1}\bar{1}0$] & [$1\bar{1}$] & [$\bar{1}1$] & [$10$] & [$\bar{1}$],$_{1}$ & [$1$],$_{1}$ & [$1$],$_{1}$ & [$0$],$_{\bar{2}}$ & $\overline{Y}$ & $\overline{Y}$ & $\overline{u_{\bar{2}}}$ & $\overline{u}$ & $\overline{u_{\bar{2}}}$ & $\overline{u}$ & $-\alpha$\\

 [$0\bar{1}1\bar{1}00$] & [$\bar{1}0$] & [$\bar{1}1$] & [$10$] & [$\bar{1}$],$_{1}$ & [$1$],$_{1}$ & [$1$],$_{1}$ & [$0$],$_{\bar{2}}$ & $\overline{Y}$ & $\overline{Y}$ & $\overline{u_{\bar{2}}}$ & $\overline{u}$ & $\overline{u_{\bar{2}}}$ & $\overline{u}$ & $-\alpha$\\

 [$00\bar{1}002$] & [$01$] & [$\bar{1}1$] & [$\bar{1}1$] & [$\bar{1}$],$_{1}$ & [$\bar{1}$],$_{1}$ & [$0$],$_{\bar{2}}$ & [$1$],$_{1}$ & $\overline{u_{\bar{2}}}$ & $\overline{u}$ & $\overline{Y}$ & $\overline{Y}$ & $\overline{u}$ & $\overline{u_{\bar{2}}}$ & $-\beta$\\

 [$01\bar{1}0\bar{1}1$] & [$1\bar{1}$] & [$\bar{1}1$] & [$\bar{1}1$] & [$\bar{1}$],$_{1}$ & [$\bar{1}$],$_{1}$ & [$0$],$_{\bar{2}}$ & [$1$],$_{1}$ & $\overline{u_{\bar{2}}}$ & $\overline{u}$ & $\overline{Y}$ & $\overline{Y}$ & $\overline{u}$ & $\overline{u_{\bar{2}}}$ & $-\beta$\\

 [$00\bar{1}001$] & [$\bar{1}0$] & [$\bar{1}1$] & [$\bar{1}1$] & [$\bar{1}$],$_{1}$ & [$\bar{1}$],$_{1}$ & [$0$],$_{\bar{2}}$ & [$1$],$_{1}$ & $\overline{u_{\bar{2}}}$ & $\overline{u}$ & $\overline{Y}$ & $\overline{Y}$ & $\overline{u}$ & $\overline{u_{\bar{2}}}$ & $-\beta$\\

 [$0\bar{1}01\bar{1}1$] & [$01$] & [$\bar{1}1$] & [$0\bar{1}$] & [$\bar{1}$],$_{1}$ & [$0$],$_{\bar{2}}$ & [$\bar{1}$],$_{1}$ & [$\bar{1}$],$_{1}$ & $\overline{u}$ & $\overline{u_{\bar{2}}}$ & $\overline{u}$ & $\overline{u_{\bar{2}}}$ & $\overline{Y}$ & $\overline{Y}$ & $-\gamma$\\

 [$0001\bar{2}0$] & [$1\bar{1}$] & [$\bar{1}1$] & [$0\bar{1}$] & [$\bar{1}$],$_{1}$ & [$0$],$_{\bar{2}}$ & [$\bar{1}$],$_{1}$ & [$\bar{1}$],$_{1}$ & $\overline{u}$ & $\overline{u_{\bar{2}}}$ & $\overline{u}$ & $\overline{u_{\bar{2}}}$ & $\overline{Y}$ & $\overline{Y}$ & $-\gamma$\\

 [$0\bar{1}01\bar{1}0$] & [$\bar{1}0$] & [$\bar{1}1$] & [$0\bar{1}$] & [$\bar{1}$],$_{1}$ & [$0$],$_{\bar{2}}$ & [$\bar{1}$],$_{1}$ & [$\bar{1}$],$_{1}$ & $\overline{u}$ & $\overline{u_{\bar{2}}}$ & $\overline{u}$ & $\overline{u_{\bar{2}}}$ & $\overline{Y}$ & $\overline{Y}$ & $-\gamma$\\
\mr
 [$\bar{1}01\bar{1}10$] & [$01$] & [$0\bar{1}$] & [$10$] & [$0$],$_{\bar{2}}$ & [$1$],$_{1}$ & [$1$],$_{1}$ & [$0$],$_{\bar{2}}$ & $d^c$ & $d^c$ & $u^c$ & $u^c_{\bar{4}}$ & $u^c$ & $u^c_{\bar{4}}$ & $\beta+\gamma$\\

 [$\bar{1}11\bar{1}0\bar{1}$] & [$1\bar{1}$] & [$0\bar{1}$] & [$10$] & [$0$],$_{\bar{2}}$ & [$1$],$_{1}$ & [$1$],$_{1}$ & [$0$],$_{\bar{2}}$ & $d^c$ & $d^c$ & $u^c$ & $u^c_{\bar{4}}$ & $u^c$ & $u^c_{\bar{4}}$ & $\beta+\gamma$\\

 [$\bar{1}01\bar{1}1\bar{1}$] & [$\bar{1}0$] & [$0\bar{1}$] & [$10$] & [$0$],$_{\bar{2}}$ & [$1$],$_{1}$ & [$1$],$_{1}$ & [$0$],$_{\bar{2}}$ & $d^c$ & $d^c$ & $u^c$ & $u^c_{\bar{4}}$ & $u^c$ & $u^c_{\bar{4}}$ & $\beta+\gamma$\\

 [$\bar{1}1\bar{1}011$] & [$01$] & [$0\bar{1}$] & [$\bar{1}1$] & [$0$],$_{\bar{2}}$ & [$\bar{1}$],$_{1}$ & [$0$],$_{\bar{2}}$ & [$1$],$_{1}$ & $u^c$ & $u^c_{\bar{4}}$ & $d^c$ & $d^c$ & $u^c_{\bar{4}}$ & $u^c$ & $\alpha+\gamma$\\

 [$\bar{1}2\bar{1}000$] & [$1\bar{1}$] & [$0\bar{1}$] & [$\bar{1}1$] & [$0$],$_{\bar{2}}$ & [$\bar{1}$],$_{1}$ & [$0$],$_{\bar{2}}$ & [$1$],$_{1}$ & $u^c$ & $u^c_{\bar{4}}$ & $d^c$ & $d^c$ & $u^c_{\bar{4}}$ & $u^c$ & $\alpha+\gamma$\\

 [$\bar{1}1\bar{1}010$] & [$\bar{1}0$] & [$0\bar{1}$] & [$\bar{1}1$] & [$0$],$_{\bar{2}}$ & [$\bar{1}$],$_{1}$ & [$0$],$_{\bar{2}}$ & [$1$],$_{1}$ & $u^c$ & $u^c_{\bar{4}}$ & $d^c$ & $d^c$ & $u^c_{\bar{4}}$ & $u^c$ & $\alpha+\gamma$\\

 [$\bar{1}00100$] & [$01$] & [$0\bar{1}$] & [$0\bar{1}$] & [$0$],$_{\bar{2}}$ & [$0$],$_{\bar{2}}$ & [$\bar{1}$],$_{1}$ & [$\bar{1}$],$_{1}$ & $u^c_{\bar{4}}$ & $u^c$ & $u^c_{\bar{4}}$ & $u^c$ & $d^c$ & $d^c$ & $\alpha+\beta$\\

 [$\bar{1}101\bar{1}\bar{1}$] & [$1\bar{1}$] & [$0\bar{1}$] & [$0\bar{1}$] & [$0$],$_{\bar{2}}$ & [$0$],$_{\bar{2}}$ & [$\bar{1}$],$_{1}$ & [$\bar{1}$],$_{1}$ & $u^c_{\bar{4}}$ & $u^c$ & $u^c_{\bar{4}}$ & $u^c$ & $d^c$ & $d^c$ & $\alpha+\beta$\\

 [$\bar{1}0010\bar{1}$] & [$\bar{1}0$] & [$0\bar{1}$] & [$0\bar{1}$] & [$0$],$_{\bar{2}}$ & [$0$],$_{\bar{2}}$ & [$\bar{1}$],$_{1}$ & [$\bar{1}$],$_{1}$ & $u^c_{\bar{4}}$ & $u^c$ & $u^c_{\bar{4}}$ & $u^c$ & $d^c$ & $d^c$ & $\alpha+\beta$\\
\mr
{\bf 78}  &  {\bf 3} & $\bf\overline{3}$ & $\bf\overline{3}$\\
\ns\mr
 [$100\bar{1}01$] & [$10$] & [$01$] & [$01$] & [$0$],$_{2}$ & [$0$],$_{2}$ & [$1$],$_{\bar{1}}$ & [$1$],$_{\bar{1}}$ & $\overline{u^c_{\bar{4}}}$ & $\overline{u^c}$ & $\overline{u^c_{\bar{4}}}$ & $\overline{u^c}$ & $\overline{d^c}$ & $\overline{d^c}$ & $-\alpha-\beta$\\

 [$1\bar{1}0\bar{1}11$] & [$\bar{1}1$] & [$01$] & [$01$] & [$0$],$_{2}$ & [$0$],$_{2}$ & [$1$],$_{\bar{1}}$ & [$1$],$_{\bar{1}}$ & $\overline{u^c_{\bar{4}}}$ & $\overline{u^c}$ & $\overline{u^c_{\bar{4}}}$ & $\overline{u^c}$ & $\overline{d^c}$ & $\overline{d^c}$ & $-\alpha-\beta$\\

 [$100\bar{1}00$] & [$0\bar{1}$] & [$01$] & [$01$] & [$0$],$_{2}$ & [$0$],$_{2}$ & [$1$],$_{\bar{1}}$ & [$1$],$_{\bar{1}}$ & $\overline{u^c_{\bar{4}}}$ & $\overline{u^c}$ & $\overline{u^c_{\bar{4}}}$ & $\overline{u^c}$ & $\overline{d^c}$ & $\overline{d^c}$ & $-\alpha-\beta$\\

 [$1\bar{1}10\bar{1}0$] & [$10$] & [$01$] & [$1\bar{1}$] & [$0$],$_{2}$ & [$1$],$_{\bar{1}}$ & [$0$],$_{2}$ & [$\bar{1}$],$_{\bar{1}}$ & $\overline{u^c}$ & $\overline{u^c_{\bar{4}}}$ & $\overline{d^c}$ & $\overline{d^c}$ & $\overline{u^c_{\bar{4}}}$ & $\overline{u^c}$ & $-\alpha-\gamma$\\

 [$1\bar{2}1000$] & [$\bar{1}1$] & [$01$] & [$1\bar{1}$] & [$0$],$_{2}$ & [$1$],$_{\bar{1}}$ & [$0$],$_{2}$ & [$\bar{1}$],$_{\bar{1}}$ & $\overline{u^c}$ & $\overline{u^c_{\bar{4}}}$ & $\overline{d^c}$ & $\overline{d^c}$ & $\overline{u^c_{\bar{4}}}$ & $\overline{u^c}$ & $-\alpha-\gamma$\\

 [$1\bar{1}10\bar{1}\bar{1}$] & [$0\bar{1}$] & [$01$] & [$1\bar{1}$] & [$0$],$_{2}$ & [$1$],$_{\bar{1}}$ & [$0$],$_{2}$ & [$\bar{1}$],$_{\bar{1}}$ & $\overline{u^c}$ & $\overline{u^c_{\bar{4}}}$ & $\overline{d^c}$ & $\overline{d^c}$ & $\overline{u^c_{\bar{4}}}$ & $\overline{u^c}$ & $-\alpha-\gamma$\\

 [$10\bar{1}1\bar{1}1$] & [$10$] & [$01$] & [$\bar{1}0$] & [$0$],$_{2}$ & [$\bar{1}$],$_{\bar{1}}$ & [$\bar{1}$],$_{\bar{1}}$ & [$0$],$_{2}$ & $\overline{d^c}$ & $\overline{d^c}$ & $\overline{u^c}$ & $\overline{u^c_{\bar{4}}}$ & $\overline{u^c}$ & $\overline{u^c_{\bar{4}}}$ & $-\beta-\gamma$\\

 [$1\bar{1}\bar{1}101$] & [$\bar{1}1$] & [$01$] & [$\bar{1}0$] & [$0$],$_{2}$ & [$\bar{1}$],$_{\bar{1}}$ & [$\bar{1}$],$_{\bar{1}}$ & [$0$],$_{2}$ & $\overline{d^c}$ & $\overline{d^c}$ & $\overline{u^c}$ & $\overline{u^c_{\bar{4}}}$ & $\overline{u^c}$ & $\overline{u^c_{\bar{4}}}$ & $-\beta-\gamma$\\

 [$10\bar{1}1\bar{1}0$] & [$0\bar{1}$] & [$01$] & [$\bar{1}0$] & [$0$],$_{2}$ & [$\bar{1}$],$_{\bar{1}}$ & [$\bar{1}$],$_{\bar{1}}$ & [$0$],$_{2}$ & $\overline{d^c}$ & $\overline{d^c}$ & $\overline{u^c}$ & $\overline{u^c_{\bar{4}}}$ & $\overline{u^c}$ & $\overline{u^c_{\bar{4}}}$ & $-\beta-\gamma$\\

\mr
 [$010\bar{1}10$] & [$10$] & [$1\bar{1}$] & [$01$] & [$1$],$_{\bar{1}}$ & [$0$],$_{2}$ & [$1$],$_{\bar{1}}$ & [$1$],$_{\bar{1}}$ & $u$ & $u_{\bar{2}}$ & $u$ & $u_{\bar{2}}$ & $Y$ & $Y$ & $\gamma$\\

 [$000\bar{1}20$] & [$\bar{1}1$] & [$1\bar{1}$] & [$01$] & [$1$],$_{\bar{1}}$ & [$0$],$_{2}$ & [$1$],$_{\bar{1}}$ & [$1$],$_{\bar{1}}$ & $u$ & $u_{\bar{2}}$ & $u$ & $u_{\bar{2}}$ & $Y$ & $Y$ & $\gamma$\\

 [$010\bar{1}1\bar{1}$] & [$0\bar{1}$] & [$1\bar{1}$] & [$01$] & [$1$],$_{\bar{1}}$ & [$0$],$_{2}$ & [$1$],$_{\bar{1}}$ & [$1$],$_{\bar{1}}$ & $u$ & $u_{\bar{2}}$ & $u$ & $u_{\bar{2}}$ & $Y$ & $Y$ & $\gamma$\\

 [$00100\bar{1}$] & [$10$] & [$1\bar{1}$] & [$1\bar{1}$] & [$1$],$_{\bar{1}}$ & [$1$],$_{\bar{1}}$ & [$0$],$_{2}$ & [$\bar{1}$],$_{\bar{1}}$ & $u_{\bar{2}}$ & $u$ & $Y$ & $Y$ & $u$ & $u_{\bar{2}}$ & $\beta$\\

 [$0\bar{1}101\bar{1}$] & [$\bar{1}1$] & [$1\bar{1}$] & [$1\bar{1}$] & [$1$],$_{\bar{1}}$ & [$1$],$_{\bar{1}}$ & [$0$],$_{2}$ & [$\bar{1}$],$_{\bar{1}}$ & $u_{\bar{2}}$ & $u$ & $Y$ & $Y$ & $u$ & $u_{\bar{2}}$ & $\beta$\\

 [$00100\bar{2}$] & [$0\bar{1}$] & [$1\bar{1}$] & [$1\bar{1}$] & [$1$],$_{\bar{1}}$ & [$1$],$_{\bar{1}}$ & [$0$],$_{2}$ & [$\bar{1}$],$_{\bar{1}}$ & $u_{\bar{2}}$ & $u$ & $Y$ & $Y$ & $u$ & $u_{\bar{2}}$ & $\beta$\\

 [$01\bar{1}100$] & [$10$] & [$1\bar{1}$] & [$\bar{1}0$] & [$1$],$_{\bar{1}}$ & [$\bar{1}$],$_{\bar{1}}$ & [$\bar{1}$],$_{\bar{1}}$ & [$0$],$_{2}$ & $Y$ & $Y$ & $u_{\bar{2}}$ & $u$ & $u_{\bar{2}}$ & $u$ & $\alpha$\\

 [$00\bar{1}110$] & [$\bar{1}1$] & [$1\bar{1}$] & [$\bar{1}0$] & [$1$],$_{\bar{1}}$ & [$\bar{1}$],$_{\bar{1}}$ & [$\bar{1}$],$_{\bar{1}}$ & [$0$],$_{2}$ & $Y$ & $Y$ & $u_{\bar{2}}$ & $u$ & $u_{\bar{2}}$ & $u$ & $\alpha$\\

 [$01\bar{1}10\bar{1}$] & [$0\bar{1}$] & [$1\bar{1}$] & [$\bar{1}0$] & [$1$],$_{\bar{1}}$ & [$\bar{1}$],$_{\bar{1}}$ & [$\bar{1}$],$_{\bar{1}}$ & [$0$],$_{2}$ & $Y$ & $Y$ & $u_{\bar{2}}$ & $u$ & $u_{\bar{2}}$ & $u$ & $\alpha$\\

 [$\bar{1}10\bar{1}01$] & [$10$] & [$\bar{1}0$] & [$01$] & [$\bar{1}$],$_{\bar{1}}$ & [$0$],$_{2}$ & [$1$],$_{\bar{1}}$ & [$1$],$_{\bar{1}}$ & $d$ & $d_{\bar{2}}$ & $d$ & $d_{\bar{2}}$ & $X$ & $X$ & $\gamma$\\

 [$\bar{1}00\bar{1}11$] & [$\bar{1}1$] & [$\bar{1}0$] & [$01$] & [$\bar{1}$],$_{\bar{1}}$ & [$0$],$_{2}$ & [$1$],$_{\bar{1}}$ & [$1$],$_{\bar{1}}$ & $d$ & $d_{\bar{2}}$ & $d$ & $d_{\bar{2}}$ & $X$ & $X$ & $\gamma$\\

 [$\bar{1}10\bar{1}00$] & [$0\bar{1}$] & [$\bar{1}0$] & [$01$] & [$\bar{1}$],$_{\bar{1}}$ & [$0$],$_{2}$ & [$1$],$_{\bar{1}}$ & [$1$],$_{\bar{1}}$ & $d$ & $d_{\bar{2}}$ & $d$ & $d_{\bar{2}}$ & $X$ & $X$ & $\gamma$\\

 [$\bar{1}010\bar{1}0$] & [$10$] & [$\bar{1}0$] & [$1\bar{1}$] & [$\bar{1}$],$_{\bar{1}}$ & [$1$],$_{\bar{1}}$ & [$0$],$_{2}$ & [$\bar{1}$],$_{\bar{1}}$ & $d_{\bar{2}}$ & $d$ & $X$ & $X$ & $d$ & $d_{\bar{2}}$ & $\beta$\\

 [$\bar{1}\bar{1}1000$] & [$\bar{1}1$] & [$\bar{1}0$] & [$1\bar{1}$] & [$\bar{1}$],$_{\bar{1}}$ & [$1$],$_{\bar{1}}$ & [$0$],$_{2}$ & [$\bar{1}$],$_{\bar{1}}$ & $d_{\bar{2}}$ & $d$ & $X$ & $X$ & $d$ & $d_{\bar{2}}$ & $\beta$\\

 [$\bar{1}010\bar{1}\bar{1}$] & [$0\bar{1}$] & [$\bar{1}0$] & [$1\bar{1}$] & [$\bar{1}$],$_{\bar{1}}$ & [$1$],$_{\bar{1}}$ & [$0$],$_{2}$ & [$\bar{1}$],$_{\bar{1}}$ & $d_{\bar{2}}$ & $d$ & $X$ & $X$ & $d$ & $d_{\bar{2}}$ & $\beta$\\

 [$\bar{1}1\bar{1}1\bar{1}1$] & [$10$] & [$\bar{1}0$] & [$\bar{1}0$] & [$\bar{1}$],$_{\bar{1}}$ & [$\bar{1}$],$_{\bar{1}}$ & [$\bar{1}$],$_{\bar{1}}$ & [$0$],$_{2}$ & $X$ & $X$ & $d_{\bar{2}}$ & $d$ & $d_{\bar{2}}$ & $d$ & $\alpha$\\

 [$\bar{1}0\bar{1}101$] & [$\bar{1}1$] & [$\bar{1}0$] & [$\bar{1}0$] & [$\bar{1}$],$_{\bar{1}}$ & [$\bar{1}$],$_{\bar{1}}$ & [$\bar{1}$],$_{\bar{1}}$ & [$0$],$_{2}$ & $X$ & $X$ & $d_{\bar{2}}$ & $d$ & $d_{\bar{2}}$ & $d$ & $\alpha$\\

 [$\bar{1}1\bar{1}1\bar{1}0$] & [$0\bar{1}$] & [$\bar{1}0$] & [$\bar{1}0$] & [$\bar{1}$],$_{\bar{1}}$ & [$\bar{1}$],$_{\bar{1}}$ & [$\bar{1}$],$_{\bar{1}}$ & [$0$],$_{2}$ & $X$ & $X$ & $d_{\bar{2}}$ & $d$ & $d_{\bar{2}}$ & $d$ & $\alpha$\\
\br
\end{tabular}
\end{flushright}
\end{table}

\end{document}